\documentclass[12pt]{iopart}
\usepackage{iopams}  
\usepackage{graphics}
\usepackage{setstack}
\usepackage{amsfonts}
\usepackage{amssymb}
\usepackage[active]{srcltx}
\usepackage[usenames,dvipsnames]{xcolor}
\usepackage{graphicx}
\usepackage{subeqnarray}
\usepackage{dcolumn}
\usepackage{pdfpages}
\usepackage{bm}
\usepackage{soul}
\definecolor{babyblue}{rgb}{0.54, 0.81, 0.94}
\definecolor{rosita}{rgb}{0.97, 0.56, 0.65}
\newcommand{\ket}[1]{\ensuremath{\left|{#1}\right\rangle}}
\newcommand{\bra}[1]{\ensuremath{\left\langle{#1}\right|}}

\newcommand{\ketbra}[2]{\ensuremath{\left|{#1}\right\rangle\!\left\langle{#2}\right|}}
\newcommand{\op}[1]{\ensuremath{\hat{\mathnormal{#1}}}}
\newcommand{\beq}{\begin{equation}}
\newcommand{\eeq}{\end{equation}}
\newcommand{\mean}[1]{\ensuremath{\langle{#1}\rangle}}
\newcommand{\Mean}[2]{\ensuremath{\langle{#1}|{#2}|{#1}\rangle}}

\begin{document}

\title[]{Dynamics of closed quantum systems under stochastic resetting}

\author{Francisco J. Sevilla and Andrea Vald\'es-Hern\'andez}
\address{Instituto de F\'isica, Universidad Nacional Aut\'onoma de M\'exico,
Apdo.\ Postal 20-364, 01000, Ciudad de M\'exico, M\'exico}
\eads{\mailto{fjsevilla@fisica.unam.mx}}
\eads{\mailto{andreavh@fisica.unam.mx}}

\vspace{10pt}
\begin{indented}
\item[]
\end{indented}

\begin{abstract}
We consider a closed quantum system subject to a stochastic resetting process. The generic expression for the resulting density operator is formulated for arbitrary resetting dynamics, fully characterised by the distribution of times between consecutive reset events. We analyse the behaviour of the state in the long-time regime, as well as the evolution of relevant quantities in the study of quantum coherence and closed- vs open-system dynamics. Our general results are complemented with examples involving paradigmatic resetting distributions, and special attention is paid to the two-level (qubit) system, in which we elucidate the effects of the renewal process on the speed of evolution toward an orthogonal state, and gain insight into the resetting applied to open systems.             
\end{abstract}

\vspace{2pc}
\noindent{\it Keywords}: Stochastic resetting, Non-equilibrium stationary state, Non-unitary dynamics, Decoherence, Mixing, Fidelity, Mean orthogonality time.

%

\section{Introduction}

The sudden restarting of a process to a predetermined state at random times has acquired a growing interest since the seminal paper of Evans and Majumdar \cite{EvansPRL2011}, who studied the resetting to a fixed position of a one-dimensional diffusion process. From then on, it has been shown that stochastic resetting has a profound impact on the properties of systems evolving under a stochastic and deterministic dynamics \cite{FuchsEPL2016,ReuveniPRL2016,PalPRL2017,EvansJPhysA2020,GuptaPRL2020,MagoniPRR2020,RayChaos2021,SarkarChaos2022}.
In the case of closed quantum systems, whose states evolve under unitary quantum dynamics, Mukherjee et al.~\cite{MukherjeePRB2018} studied the non-equilibrium dynamics induced on some integrable and nonintegrable systems as a result of their resetting to an initial fix state, performed at random exponentially distributed times. Particular attention has been paid to the analysis of the stationary state generated as a result of the resetting in both, closed and open quantum systems \cite{RosePRE2018}, and for generalised resetting protocols in a quantum Ising chain in transverse field \cite{PerfettoPRB2021}. More recently \cite{TurkeshiPRB2022}, different entanglement scaling has been observed to arise in an Ising chain under the effects of a measurement protocol as stochastic resetting process on the ballistic propagation of well-defined quasiparticles. 

To our knowledge, the stochastic resetting process considered so far
in the quantum realm, is restricted to the case in which the time between consecutive resetting events is exponentially distributed, and the characteristic effects of stochastic resetting on central properties of quantum dynamics  (such as coherence, purity and quantum fidelity) have not been analysed in the more general scenario, which involves arbitrary states subject to a resetting process in which the time between consecutive events follows a generic distribution. The present paper intends to fill this gap. 


We formulate the renewal equation for the density operator of a quantum system subject to an arbitrary stochastic resetting process, which is fully characterised by the probability distribution of the time intervals between consecutive resetting events. We investigate the relation between the resetting distribution and the structure of the steady state (when exists), attained at sufficiently long times. Further, we show that the resetting process deeply influences the otherwise unitary dynamics of closed systems, by analysing its main effects on relevant quantities such as the coherence, the purity, and the fidelity with respect to the non-reset state. The paradigmatic two-level (qubit) system is particularly studied, allowing for an analysis of the delay induced by the resetting on the speed of evolution of the system toward an orthogonal state, and also for an application of the renewal equation to open systems.

\section{Stochastic resetting process}

Consider the time evolution of a quantum system, either isolated or open, whose initial state is given by the density matrix $\op{\rho}(0)$.  
In
the most general case the evolution is generically described by a dynamical map (or quantum channel) represented by the superoperator $\mathcal{E}(t)$ such that $\hat \rho(t)=\mathcal{E}(t)\hat \rho(0)$ \cite{Rivas2012open}; then we allow this evolution to be interrupted by stochastic events that reset the system's state to the initial one, and then let it evolve again under $\mathcal E$ till the next reset event. The evolution of the system is thus dictated by the dynamical map between any pair of consecutive reset events; if these occur at times $s$ and $s^{\prime}>s$, then for
$t\in(s,s^{\prime})$
\begin{equation}
\op{\rho}(s<t<s^{\prime})=\mathcal{E}(t-s)\hat \rho(s)=\mathcal{E}(t-s)\hat \rho(0)=\hat \rho(t-s).
\end{equation}
Thus at time $t$, one element of the statistical ensemble generated by stochastic resetting is given by $\op{\rho}(t-s)$, its probability of occurrence being the probability that a reset occurred at time $s$ without occurring another one in the interval $t-s$, i.e.,
\begin{equation}
\mathcal{T}(s)\mathcal{T}_{0}(t-s)ds,
\end{equation}
where $\mathcal{T}(t)$ denotes the probability density that a reset event occurs at time $t$, while $\mathcal{T}_{0}(T)$ stands for the probability that no reset event occurs during the time interval of duration $T$. At time $t$ we therefore consider the statistical mixture defined by 
\begin{equation}\label{StateResettingPart}
\int_{0}^{t}ds\, \mathcal{T}(s)\mathcal{T}_{0}(t-s)\op{\rho}(t-s).
\end{equation}
One more element has to be included in the ensemble, this corresponds to the state that has evolved under $\mathcal{E}$ from $\op{\rho}(0)$ to $\op{\rho}(t)$ without being interrupted by any reset event
, and is given by $\op{\rho}(t)$ times  $\mathcal{T}_{0}(t)$. The density operator under the stochastic resetting process, denoted as $\op{\rho}_{\rm sr}$, is thus given by
\begin{equation}\label{DensityOP}
 \op{\rho}_{\rm sr}(t)=\mathcal{T}_{0}(t)\op{\rho}(t)+\int_{0}^{t}ds\, \mathcal{T}(s)\mathcal{T}_{0}(t-s)\op{\rho}(t-s).
\end{equation}
This is the renewal equation that gives the system's state at time $t$ (similar renewal equations have been formulated to describe classical-diffusive transport \cite{NagarPRE2016,ChechkinPRL2018,Maso-PuigdellosasPRE2019,MasoliverPRE2019}). 

The normalisation condition of the density matrix, $\Tr \op{\rho}_{\rm sr}(t)=1$, is consistent with the normalisation of the probabilities of the ensemble, namely
\beq\label{traza1}
1=\mathcal{T}_{0}(t)+\int_{0}^{t}ds\, \mathcal{T}(s)\mathcal{T}_{0}(t-s),
\eeq
as can be easily checked by taking the trace on both sides in \Eref{DensityOP}. 
\subsection{\label{subsect:ParticularProcesses}
The distribution of times between consecutive resets}

The renewal process defined by the stochastic resetting just described can be fully characterised in terms of 
the distribution $\mathcal{T}_1(T)$ of the time intervals of duration $T$ between two consecutive reset events, which in the simplest case are independent and identically distributed random variables. Notice that $\mathcal{T}_1(t)$ coincides with the probability density that the \emph{first} reset event occurs at time $t$, reason for choosing the notation $\mathcal{T}_{1}(t)$. Since at time $t$ either no reset has occurred, or a first reset has taken place, it holds that
\begin{equation}\label{T0}
\mathcal{T}_0(t)=1-\int_{0}^{t}ds\ \mathcal{T}_1(s).    
\end{equation}

The connection between $\mathcal{T}_1$ and $\mathcal{T}$ is less straightforward. To establish it we first realize that if $\mathcal{T}_n(t)$ denotes the probability density that the $n$-th reset event ($n\ge1$) occurs at time $t$, then  $\mathcal{T}(t)=\sum_{n=1}^{\infty}\mathcal{T}_{n}(t)$. Further, $\mathcal{T}_{n}(t)$ can be obtained recursively from  
$\mathcal{T}_n(t)=\int_{0}^{t}ds\, \mathcal{T}_{n-1}(t-s)\mathcal{T}_{1}(s)$, whose Laplace transform gives $\widetilde{\mathcal{T}}_n(\epsilon)=\widetilde{\mathcal{T}}_{n-1}(\epsilon)\widetilde{\mathcal{T}}_1(\epsilon)$. Successive application of this result gives
$\widetilde{\mathcal{T}}_{n}(\epsilon)=\biggl[\widetilde{\mathcal{T}}_{1}(\epsilon)\biggr]^{n}$, thus leading to
\begin{equation}\label{TfromT1}
\widetilde{\mathcal{T}}(\epsilon)=\sum_{n=1}^{\infty}\biggl[\widetilde{\mathcal{T}}_{1}(\epsilon)\biggr]^{n}.
\end{equation}
Computation of the inverse Laplace transform of this result, when possible, gives us the wanted relation between $\mathcal{T}_1$ and $\mathcal{T}$. In some cases, the sum in \Eref{TfromT1} can be formally evaluated and gives
\begin{equation}\label{TfromT1bis}
\widetilde{\mathcal{T}}(\epsilon)=\frac{\widetilde{\mathcal{T}}_{1}(\epsilon)}{1-\widetilde{\mathcal{T}}_{1}(\epsilon)}.
\end{equation}
Therefore the specification of $\mathcal{T}_{1}(t)$ completely determines the resetting process, via the determination of $\mathcal{T}_0$ and $\mathcal{T}$.

We are interested in resetting processes consistent with the physically-motivated requirement that $\mathcal{T}_{0}(t)$ vanishes asymptotically and monotonically with time.
This last statement is expressed in terms of the Laplace transform of $\mathcal{T}_{0}(t)$, $\widetilde{\mathcal{T}}_{0}(\epsilon)$, as
\beq\label{limT0}
\lim_{\epsilon\rightarrow0}\epsilon\widetilde{\mathcal{T}}_{0}(\epsilon)=0.
\eeq
 Such condition requires an asymptotic behaviour for $\mathcal{T}_{1}(t)$ that is elucidated by means of Eq. (\ref{T0}), from which we have that $\widetilde{\mathcal{T}}_{0}(\epsilon)=[1-\widetilde{\mathcal{T}}_{1}(\epsilon)]/\epsilon$ and therefore $\lim_{\epsilon\rightarrow0}\widetilde{\mathcal{T}}_{1}(\epsilon)=1^{-}$, indicating that $\widetilde{\mathcal{T}}_{0}(\epsilon)$ tends to 1 from the left as $\epsilon$ goes to 0. The class of distributions $\mathcal{T}_{1}(t)$ that comply with this requirement is indeed large, however two subclasses
can be highlighted for  $\mathcal{T}_{1}$ characterised by two real parameters $\alpha,\, \beta>0$, and correspond to distributions for which, given $\epsilon/\alpha\ll1$, can be approximated as: i) $\widetilde{\mathcal{T}}_{1}(\epsilon)=1-\beta(\epsilon/\alpha)+{\rm O}(\epsilon^{2})$, and ii) $\widetilde{\mathcal{T}}_{1}(\epsilon)=1-(\epsilon/\alpha)^{\beta}+{\rm O}(\epsilon^{2})$.


To the first subclass belong distributions whose Laplace transform has the structure
\begin{equation}\label{T1subclass1}
\widetilde{\mathcal{T}}_{1}(\epsilon)=\frac{\alpha^{\beta}}{(\epsilon+\alpha)^{\beta}}.    
\end{equation}
This is the case of the \emph{gamma distribution}, which reduces to the \emph{exponential distribution} for $\beta=1$. Distributions for which
\begin{equation}\label{T1subclass2}
\widetilde{\mathcal{T}}_{1}(\epsilon)=e^{-(\epsilon/\alpha)^{\beta}}
\end{equation}
pertain to the second subclass and include, for $\beta=1/2$, the \emph{L\'evy-Smirnoff distribution}. In the following we introduce these distributions in some detail.

\begin{figure}
    \centering
    \includegraphics[width=\textwidth,trim=0 15 0 0]{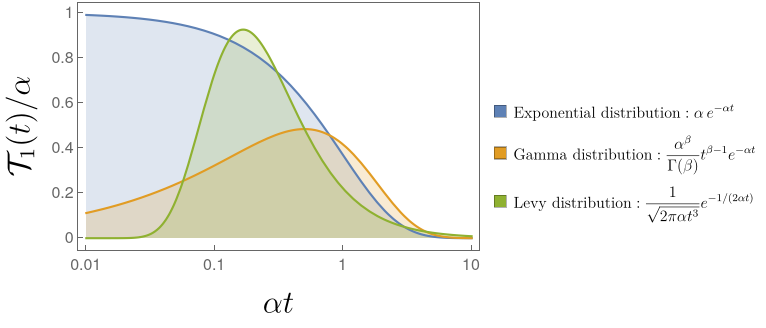}
    \caption{Dimensionless probability distribution $\mathcal{T}_1(t)/\alpha$ between consecutive resetting events as a function of the dimensionless time $\alpha t$. 
    Short-time intervals are more frequently sampled for the exponential distribution (blue curve)
     than for the gamma distribution if $\beta>1$ (orange curve, with $\beta=3/2$), while these are seldom sampled for the Lévy-Smirnoff distribution (green curve).}
    \label{fig:Ejemplos}
\end{figure}

The exponential distribution has been previously considered in the study of stochastic resetting in quantum dynamics \cite{MukherjeePRB2018}
. It corresponds to 
\begin{equation}\label{ExpDist}
\mathcal{T}_{1}(t)=\alpha e^{-\alpha t}    
\end{equation}
with $\alpha> 0$ being the resetting rate, which in this case coincides with the inverse of the mean time between resetting events, i.e. $\langle t\rangle_{\mathcal{T}_1}\equiv\int_{0}^{\infty}\, t\, \mathcal{T}_{1}(t)\,dt=\alpha^{-1}$.  Thus straightforwardly we have $\mathcal{T}(t)=\alpha$, and $\mathcal{T}_{0}(t)=e^{-\alpha t}$. This distribution gives more weight to the smaller resetting times as is shown by the blue curve in Fig. \ref{fig:Ejemplos}. 

The gamma distribution
generalises the exponential one in that the modulation of the small times between resetting events is given by a power law, i.e.
\begin{equation}\label{GammaDist}
\mathcal{T}_{1}(t)=\frac{\alpha^{\beta}}{\Gamma(\beta)}t^{\beta-1}e^{-\alpha t},    
\end{equation}
where $\alpha,\beta>0$, and $\Gamma(\sigma)=\int_{0}^{\infty}dx\, e^{-x}x^{\sigma-1}$ denotes the standard gamma function. The main characteristics of the gamma distribution are that 
it vanishes as a power law for small times, it has a maximum at a finite characteristic time $t_{\rm max}=(\beta-1)\alpha^{-1}=(1-\beta^{-1})\langle t\rangle_{\mathcal{T}_{1}}$ if $\beta>1$ (no such finite characteristic time exists when $\beta\le1$ since $t_{\rm max}=0$), and decays exponentially for large times as is shown (for $\beta=3/2$) by the orange curve in Fig. \ref{fig:Ejemplos}. In this case $\mathcal{T}_{0}(t)=\Gamma(\beta,\alpha t)/\Gamma(\beta)$ where $\Gamma(\sigma,z)=\int_{z}^{\infty}dx\, e^{-x}x^{\sigma-1}$ stands for the upper incomplete gamma function, and  $\mathcal{T}(t)=\alpha e^{-\alpha t}(\alpha t)^{\beta}\, {\rm E}_{\beta,\beta}\bigl(\alpha^{\beta}t^{\beta}\bigr)$, with ${\rm E}_{\mu,\nu}(z)=\sum_{n=0}^{\infty}\frac{z^{n}}{\Gamma(\mu n+\nu)}$ the Mittag-Leffler function in two parameters $\mu,\nu>0$.

The L\'evy-Smirnoff distribution corresponds to
\begin{equation}\label{LevyDist}
\mathcal{T}_{1}(t)=\frac{1}{\sqrt{2\pi\alpha t^{3}}}e^{-\frac{1}{2\alpha t}}
\end{equation}
with $\alpha> 0$. At a strike contrast with the previous distributions, this one does not posses finite (integer) moments; in particular no finite mean time between resetting events exists. 
This prevents us from associating the parameter $\alpha$ with the distribution's mean, yet $\mathcal{T}_1(t)$ has a maximum at  $t_{\rm max}=(3\alpha)^{-1}$, allowing to associate $\alpha$ with the distribution's mode. 
The function (\ref{LevyDist}) vanishes exponentially fast for small times, while it decays slowly as a power law with exponent $3/2$ for sufficiently large times, as shown by the green curve in Figure \ref{fig:Ejemplos}. Direct calculation from Eq. (\ref{T0}) leads to $\mathcal{T}_{0}(t)={\rm erf}\Bigl(\frac{1}{\sqrt{2\alpha t}}\Bigr)$, where ${\rm erf}(x)$ denotes the error function. Finally, to our knowledge, $\mathcal{T}(t)$ cannot be obtained from $\mathcal{T}_{1}(t)$ in an straightforward manner; however it is possible in the domain of the Laplace transform by use of Eq. (\ref{TfromT1}), which gives
\begin{equation}\label{TLevyLaplace}
\widetilde{\mathcal{T}}(\epsilon)=\frac{1}{e^{\sqrt{2\epsilon/\alpha}}-1},   
\end{equation}
where we have used that  $\widetilde{\mathcal{T}}_{1}(\epsilon)=e^{-\sqrt{2\epsilon/\alpha}}$. The large-time ($\alpha t\gg1$) behaviour of $\mathcal{T}(t)$ follows from \eref{TLevyLaplace} by approximating it by $\widetilde{\mathcal{T}}(\epsilon)\approx\sqrt{\frac{\alpha}{2}}\epsilon^{-1/2}$ for $\epsilon/\alpha\ll1$, which leads to $\mathcal{T}(t)\sim\frac{\alpha}{2}\sqrt{\frac{1}{\pi t}}$. Similarly, in the short-time regime ($\alpha t\ll1$), 
$\mathcal{T}(t)$ decays exponentially as  $\mathcal{T}_{1}(t)$ does, since in this case we have $\epsilon/\alpha\gg1$ and thereby $\widetilde{\mathcal{T}}(\epsilon)\sim e^{-\sqrt{2\epsilon/\alpha}}$.

\section{Generic effects of stochastic resetting on closed systems}

Throughout this section we will explore the reset state in the long-time regime, and the main effects of the resetting process on 
quantities that are relevant in the study of quantum information and the dynamics of quantum states, namely, the coherence, the purity and the fidelity. Unless otherwise stated, we will focus on 
energy-preserving maps represented by unitary transformations
$\hat U=e^{-i\hat H t/\hbar}$ generated by a time-independent Hamiltonian $\hat{H}$, so 
\begin{equation}\label{RhoUnitary}
\op{\rho}(t)=\op{U}(t)\op{\rho}(0)\op{U}^{\dagger}(t).
\end{equation}
The eigenvectors of $\hat H$, their corresponding eigenvalues and the associated transition frequencies are denoted as $\{\ket{n}\}$, $\{E_{n}\}$, and $\omega_{nm}=(E_n-E_m)/\hbar$, respectively.

The matrix elements of $\op{\rho}_{\rm sr}(t)$ in the energy representation read 
\begin{equation}\label{rhomatrix}
 \langle n|\op{\rho}_{\rm sr}(t)|m\rangle=\langle n|\hat\rho(t)|m\rangle I_{nm}(t)=\langle n|\hat\rho(0)|m\rangle e^{-i\omega_{nm}t} I_{nm}(t)
\end{equation}
where
\beq\label{Inm}
I_{nm}(t)=\mathcal{T}_{0}(t)+
 \int_0^t ds\,  e^{i\omega_{nm}s}
 \mathcal {T}(s)\mathcal{T}_0(t-s),
\eeq
so
\begin{equation}
\bigl|I_{nm}(t)\bigr|\leq\mathcal{T}_{0}(t)+\int_0^t ds\,  \mathcal {T}(s)\mathcal{T}_0(t-s)=1.\end{equation}
By the normalization condition (\ref{traza1}) we have $I_{nm}=1$ whenever $\omega_{nm}=0$, whence the diagonal and the off-diagonal elements involving degenerate states ($\langle n|\hat\rho|n^{\prime}\rangle$ with $E_n=E_{n^{\prime}}$), are unaffected by stochastic resetting. In particular, initial states that are diagonal in the basis $\{\ket{n}\}$ are robust against the resetting (as expected since they are also invariant under the unitary evolution). In what follows we assume that there are no energy degeneracies, so $\omega_{nm}=0$ amounts to $n=m$, and from Eq. (\ref{rhomatrix}) we are led to 
\beq \label{rhosrexpansion}
\op{\rho}_{\rm sr}(t)=
\op{\rho}_{\rm D}+\sum_{n\neq m} e^{-i\omega_{nm}t}I_{nm}(t) \rho_{nm}(0)\ket{n}\!\bra{m},
\eeq
with $\op{\rho}_{\rm D}\equiv\sum_{n} \rho_{nn}(0)\ket{n}\!\bra{n}$ and $\rho_{nm}\equiv\langle n|\hat \rho|m\rangle$.


\subsection{The transition to the equilibrate state}\label{equilibrate_state}

A most interesting feature of the stochastic resetting process is that it may drive the system to an out-of-equilibrium stationary state, both in classical \cite{EvansJPhysA2020} and quantum domains \cite{RosePRE2018,PerfettoPRB2021}; in particular the attainment of a steady state in closed quantum systems has been of general interest \cite{TasakiPRL1998,PolkovnikovRMP2011}. Despite its relevance, the study of the emergence of stationary states under stochastic resetting has been circumscribed to the exponential distribution of time-intervals between consecutive resetting events (the gamma distribution has been considered in the analysis of classical diffusion \cite{RadiceJPhysA2022}). Here we overcome this limitation by considering arbitrary, however well-behaved, resetting dynamics.

We are interested in finding general conditions on $\mathcal{T}_{1}(t)$ that guarantee the existence of
an equilibrate state\footnote{Equilibration here refers to the evolution of the system towards some particular state (typically mixed), or a state close to it, in which the system remains for almost all times. Such equilibrium state may depend on the initial state.} $\op{\rho}_{\rm sr,eq}$, defined as
\beq
\op{\rho}_{\rm sr,eq}\equiv\lim_{t\rightarrow\infty}\op{\rho}_{\rm sr}(t).
\eeq
With this aim we resort to the Laplace transform of $\op{\rho}_{\rm sr}(t)$, namely $\widetilde{\rho}_{\rm sr}(\epsilon)$, and assume that the  conditions under which the \emph{final value theorem} holds are satisfied, so we can write
\beq \label{rhoEqDef}
\op{\rho}_{\rm sr,eq}\equiv\lim_{t\rightarrow\infty}\op{\rho}_{\rm sr}(t)=\lim_{\epsilon\rightarrow0}\epsilon\widetilde{\rho}_{\rm sr}(\epsilon).
\eeq
Notice that the operator $\epsilon\widetilde{\rho}_{\rm sr}(\epsilon)$ is a legitimate density operator (satisfying the Hermiticity, normalisation and semi positive-definite conditions) provided  $\epsilon\geq 0$, whence we will restrict the Laplace transform considering only a non-negative Laplace variable. Taking the Laplace transform of (\ref{rhosrexpansion}) and multiplying by $\epsilon$ results in 
\begin{equation}\label{rhosrLaplace}
\epsilon\widetilde{\rho}_{\rm sr}(\epsilon)=
\op{\rho}_{\rm D}+\sum_{n\neq m}\epsilon\widetilde{I}_{nm}(\epsilon+i\omega_{nm})\rho_{nm}(0)\ketbra{n}{m}.
\end{equation}
Assuming that the limits in Eq. (\ref{rhoEqDef}) exists, we arrive at
\begin{equation}\label{sreq}
\op{\rho}_{\rm sr,eq}=
\op{\rho}_{\rm D}+\sum_{n\neq m}\biggl[\lim_{\epsilon\rightarrow0}\epsilon\widetilde{I}_{nm}(\epsilon+i\omega_{nm})\biggr]\rho_{nm}(0)\ketbra{n}{m}.
\end{equation}
To determine the second term (off-diagonal elements) in the last expression we resort to the definition \eref{Inm} and write
\begin{equation}\label{limItilde}
\lim_{\epsilon\rightarrow0}\epsilon\, \widetilde{I}_{nm}(\epsilon+i\omega_{nm})=\lim_{\epsilon\rightarrow0}\Bigl[\epsilon\,\widetilde{\mathcal{T}}_{0}(\epsilon+i\omega_{nm})+\epsilon\,\widetilde{\mathcal{T}}_{0}(\epsilon+i\omega_{nm})\widetilde{\mathcal{T}}(\epsilon)\Bigr].    
\end{equation}
In line with the condition (\ref{limT0}) it is natural to expect that
$\lim_{\epsilon\rightarrow0}\widetilde{\mathcal{T}}_{0}(\epsilon+i\omega_{nm})$ exists and is given by $\widetilde{\mathcal{T}}_{0}(i\omega_{nm})$, thus determining (\ref{limItilde})
reduces to calculate
\beq\label{Lim}
\mathsf{L}\equiv\lim_{\epsilon\rightarrow0}\epsilon\widetilde{\mathcal{T}}(\epsilon).
\eeq
Notice that when the relation \eref{TfromT1bis} holds we have 
$\mathsf L=\lim_{\epsilon\rightarrow0}\frac{\epsilon\widetilde{\mathcal{T}}_{1}(\epsilon)}{1-\widetilde{\mathcal{T}}_{1}(\epsilon)}$. With these considerations (and assuming that $\mathsf L$ exists, in which case it is real) the equilibrate state (\ref{sreq}) is

\begin{equation}\label{eqState}
\op{\rho}_{\rm sr,eq}=
\op{\rho}_{\rm D}+{\mathsf L}\sum_{n\neq m}\widetilde{\mathcal{T}}_{0}(i\omega_{nm})\rho_{nm}(0)\ketbra{n}{m}, \end{equation}
and we can identify
\beq\label{lim}
\lim_{t\rightarrow\infty}e^{-i\omega_{nm}t}I_{nm}(t)={\mathsf L}\widetilde{\mathcal{T}}_{0}(i\omega_{nm}).
\eeq

For the distributions $\mathcal{T}_{1}(t)$ whose Laplace transform writes in the form (\ref{T1subclass1}) (gamma and exponential distributions) we have that
\beq \label{equT1}
\mathsf L=\frac{\alpha}{\beta},\quad \widetilde{\mathcal{T}}_{0}(i\omega_{nm})=\frac{1}{i\omega_{nm}}\frac{(\alpha+i\omega_{nm})^{\beta}-\alpha^{\beta}}{(\alpha+i\omega_{nm})^{\beta}}.
\eeq
The case $\beta=1$ leads to the equilibrate state reported in Ref. \cite{MukherjeePRB2018} for the exponential distribution. For $\mathcal{T}_1$ with the structure (\ref{T1subclass2}) (L\'evy-Smirnoff distribution) the limit $\mathsf L$ exits only if $0<\beta\le1$, in which case $\mathsf L=0$ and the equilibrate state is simply given by the diagonal density matrix 
\begin{equation}\label{equistate}
\op{\rho}_{\rm sr,eq}=\op{\rho}_{\rm D}=\sum_{n}\rho_{nn}(0)\ketbra{n}{n}.
\end{equation}

The results \eref{eqState} and \eref{equT1} show the strong influence of the resetting dynamics on the attainment of the stationary state.

\subsection{Coherence under the effect of stochastic resetting}\label{decoherence}

Quantum coherence lies at the core of the most paradigmatic quantum effects. It reflects the capability
of a system to exhibit interference among its possible states, and in composite systems allows for the existence of quantum correlations,
including entanglement and quantum discord \cite{AdessoJPA2016}. Quantum coherence constitutes a desirable resource in quantum information processing \cite{StreltsovRMP2017}, and the study of its quantification and dynamics has led to a growing area of research \cite{BaumgratzPRL2014,RadakrishnanPRA2019,SaxenaPRR2020,JafariPRA2020}.
 
As discussed below Eq. (\ref{Inm}), when resorting to the energy representation, $\op{\rho}_{\rm sr}(t)$ evolves with constant energy populations (diagonal elements) and only the coherence terms (off-diagonal elements) are sensitive to the resetting process. The effects of the latter on the otherwise unitary evolution are introduced via the factor $I_{nm}(t)$, whereas the combined effects of the resetting plus the unitary evolution on the initial state are encoded in the term \beq\label{Itilde}
\mathcal{I}_{nm}(t)=e^{-i\omega_{nm}t}I_{nm}(t),
\eeq
as follows from Eq. (\ref{rhomatrix}).  

When referring to coherence with respect to the eigenbasis $\{\ket{n}\}$, the \emph{free} states\footnote{This name is customary in the context of a coherence resource theory \cite{StreltsovRMP2017}.} are incoherent superpositions of states with well-defined energy, as the state $\hat\rho_D$.
For the unitary evolution (\ref{RhoUnitary}), it follows from Eq. (\ref{rhosrexpansion}) that the stochastic resetting process maps the set of free states into itself, thus corresponding to an \emph{incoherent operation}. Further, since any meaningful quantifier $\mathcal C(\hat\varrho)$ of the coherence of a generic state $\hat \varrho$ cannot increase under incoherent operations \cite{StreltsovRMP2017}, we conclude that
\beq\label{coherence}
\mathcal C[\hat\rho_{\rm sr}(t)]\leq \mathcal C[\hat\rho(0)],
\eeq
that is, the coherence does not increase as a result of the stochastic resetting.

\begin{figure}
    \centering
    \includegraphics[width=0.75\textwidth,trim=0 0 0 0]{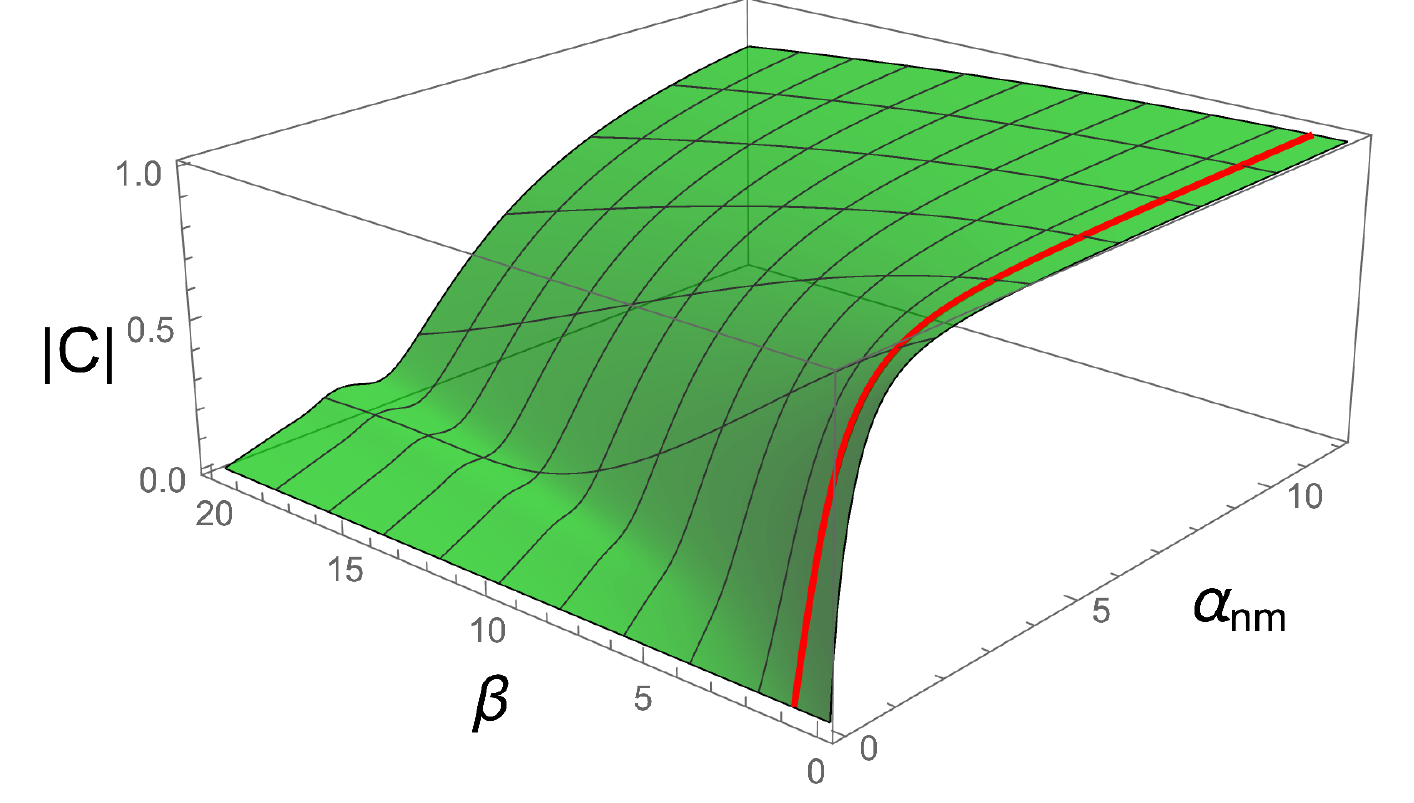}
    \caption{$C(\alpha_{nm},\beta)$ with $\alpha_{nm}=\alpha/|\omega_{nm}|$, where $\alpha$ and $\beta$ stand for the parameters of the gamma distribution. The red line corresponds to the case $\beta=1$, hence to the exponential distribution. $\alpha_{nm}\ll 1$ identifies the decoherence regime in which the state becomes a quasi-incoherent one, whereas for $\alpha_{nm}\gg 1$ the coherence $\mathcal C_{\ell_1}$ of the equilibrate state is basically the original one.}
    \label{deco}
\end{figure}

In particular, we can focus on the $\ell_1$-\emph{norm of coherence} $\mathcal{C}_{\ell_1}(\op{\varrho})=\sum_{n\neq m}|\varrho_{nm}|$ \cite{BaumgratzPRL2014} to analyse the amount of coherence at sufficiently long times, when the system approaches the equilibrate state $\hat \rho_{\rm sr,eq}$, if this exists.
In such case we may write $\langle n|\hat \rho_{\rm sr,eq}|m\rangle=\mathsf L \widetilde{\mathcal{T}}_{0}(i\omega_{nm})\rho_{nm}(0)$ (see Eqs. (\ref{rhomatrix}) and (\ref{lim})). For the gamma distribution, satisfying (\ref{T1subclass1}), we resort to Eq. (\ref{equT1}) and arrive at
\beq\label{coherence_eq}
\mathcal C_{\ell_1}(\hat \rho_{\rm sr,eq})=\sum_{n\neq m}|\langle n|\hat \rho_{\rm sr,eq}|m\rangle|=\sum_{n\neq m}| C(\alpha_{nm},\beta)||\rho_{nm}(0)|,
\eeq
with $\alpha_{nm}$ the dimensionless variable $\alpha_{nm}\equiv\alpha/|\omega_{nm}|$ and
\beq\label{coefC}
|C(\alpha_{nm},\beta)|=\Big|\frac{\alpha_{nm}}{i\beta}\Big[1-\Big(\frac{\alpha_{nm}}{\alpha_{nm}+i}\Big)^\beta\,\Big]\Big|.
\eeq
Figure \ref{deco} shows $| C(\alpha_{nm},\beta)|$, and the red curve highlights the case $\beta=1$ corresponding to the exponential distribution.
We see that for $\alpha_{nm}\ll 1$ (i.e, for $\alpha\ll \min\{|\omega_{nm}|\}$), the factor $|C(\alpha_{nm},\beta)|$ tends to zero, and $\mathcal C_{\ell_1}(\hat \rho_{\rm sr,eq})\approx 0$, so the equilibrate state is a quasi-incoherent one; we will thus refer to this as the (long-time) \emph{decoherence regime}. In the opposite extreme, in which $\alpha_{nm}\gg 1$ (meaning that $\alpha\ll \min\{|\omega_{nm}|\}$), and for values of $\beta$ of order 1, $|C(\alpha_{nm},\beta)|$ rapidly becomes $1$ and $\mathcal C_{\ell_1}(\hat \rho_{\rm sr,eq})\approx \mathcal C(\hat \rho(0))$, so the limiting coherence equals the initial one.
For the particular (exponential) case $\beta=1$, it can easily be seen that this behaviour holds not only for the equilibrate state, provided $\alpha$ is sufficiently large. Indeed, for the exponential distribution direct calculation gives   

 
\beq \label{Inmexp}
\mathcal{I}_{nm}(t)=\frac{1}{\alpha+i\omega_{nm}}\Big[\alpha+i\omega_{nm}e^{-\alpha t}e^{-i\omega_{nm}t}
\Big],
\eeq
whence for sufficiently large $\alpha$, $\mathcal{I}_{nm}$  
tends to $1$, and from Eqs. (\ref{rhomatrix}) and (\ref{Itilde}) we get $\op{\rho}_{\rm sr}\approx \hat\rho(0)$. This can be understood recalling that in the exponential case $\alpha$ represents the resetting rate, so for $1/\alpha$ much smaller than the characteristic times $1/|\omega_{nm}|$ the renewal process is so frequent that prevents the system from deviating from its initial state. Such regime corresponds thus to the Zeno effect \cite{MisraJMP77,ItanoJP2009}, here implemented via the stochastic resetting.

The above analysis shows that by an appropriate selection of $\alpha$ and $\beta$ for fixed spectrum, or rather by choosing $\hat H$ for a given pair of the gamma distribution parameters, qualitatively different regimes can be reproduced via the renewal process. In particular, the resetting performed on an otherwise closed system $S$ that evolves unitarily, can effectively simulate the non-unitary dynamics that would result if $S$ were coupled to an environment $E$ via a non-dissipative interaction (so the populations $\rho_{nn}$ remain invariant) that induces decoherence on the central system $S$. For the L\'evy-Smirnoff distribution, with equilibrate state (\ref{equistate}), the resetting process plays the role of a completely dephasing channel on $S$.

\subsection{\label{sect:Purity}Purity and Fidelity under the effect of stochastic resetting}

By construction, the resetting process maps the state $\hat \rho$ into a statistical mixture, thereby affecting its degree of mixedness while
deviating the system's state from its unitary evolution. In order to quantify such degree of mixedness and deviation, we focus on the dynamics of the \emph{purity} and the \emph{fidelity}, respectively, under the effects of stochastic resetting.

\paragraph{Purity.-} The purity $\mathcal{P}$ of $\op{\rho}_{\rm sr}(t)$ is given by
\begin{eqnarray}\label{purityA}
\mathcal{P}[\op{\rho}_{\rm sr}(t)]&\equiv&\Tr \op{\rho}_{\rm sr}^{2}(t)=\sum_{n,m}|\langle n|\op{\rho}_{\rm sr}(t)|m\rangle|^2\\
&=&\sum_{n,m}|\rho_{nm}(0)|^2|I_{nm}(t)|^2\leq \sum_{n,m}|\rho_{nm}(0)|^2=\mathcal{P}[\op{\rho}(0)]=\mathcal{P}[\op{\rho}(t)],\nonumber
\end{eqnarray}
where for writing the last line we resorted to Eq. (\ref{rhomatrix}) and used that the unitary evolution does not modify the purity of $\hat\rho(0)$. The inequality indicates that the purity reduces as a result of the stochastic resetting, and allows to identify the loss in the purity of the state as
\begin{eqnarray}\label{DeltaP}
\Delta \mathcal P(t)\equiv\mathcal{P}[\op{\rho}(0)]-\mathcal{P}[\op{\rho}_{\rm sr}(t)]=\sum_{n\neq m}|\rho_{nm}(0)|^2(1-|I_{nm}(t)|^2)\geq 0.
\end{eqnarray}

Since $\mathcal{P}[\op{\rho}(0)]\leq 1$, $\Delta \mathcal P(t)$ is bounded from above by the linear entropy associated to $\hat\rho_{\rm sr}$, namely $
S_L[\op{\rho}_{\rm sr}]\equiv 1-\Tr \op{\rho}_{\rm sr}^{2}
$. The latter is further bounded as
\begin{eqnarray}
S_L[\op{\rho}_{\rm sr}]&=&1-\sum_{n,m}|\rho_{nm}(0)|^2|I_{nm}(t)|^2\nonumber\\
&=&1-\sum_{n}[\rho_{nn}(0)]^2-\sum_{n\neq m}|\rho_{nm}(0)|^2|I_{nm}(t)|^2\nonumber\\
&\leq&1-\sum_{n}[\rho_{nn}(0)]^2=S_{L}[\{\rho_{nn}\}],
\end{eqnarray}
where $S_{L}[\{\rho_{nn}\}]$ denotes the linear entropy of the probability distribution $\{\rho_{nn}\}$. Gathering results we get
\beq\label{Deltapurbound}
\Delta \mathcal P(t)\leq S_L[\op{\rho}_{\rm sr}(t)]\leq S_{L}[\{\rho_{nn}\}].
\eeq
The first inequality in (\ref{Deltapurbound}) is saturated only for initial pure states $\hat\rho(0)=\ketbra{\psi(0)}{\psi(0)}$, and the second one only when $\sum_{n\neq m}|\rho_{nm}(0)|^2|I_{nm}(t)|^2=0$, which occurs provided $I_{nm}(t)$ vanishes for all $n\neq m$ (we rule out the possibility $\rho_{nm}(0)=0$ for $n\neq m$, since it corresponds to an incoherent initial state that is insensitive to the resetting), as when $\op{\rho}_{\rm sr,eq}$ in \Eref{eqState} becomes incoherent as was discussed in Section \ref{equilibrate_state}.
Therefore the loss in the purity tends to its maximum value $\Delta \mathcal P_{\max}=S_L[\{\rho_{nn}\}]$ whenever the initial state is pure, so
\beq\label{psi}\ket{\psi{(t)}}=\hat U\ket{\psi{(0)}}=\sum_{n}c_ne^{-iE_nt/\hbar}\ket{n},\eeq and $\hat\rho_{\rm sr}$ approximates to the incoherent state $\hat\rho_D$
(as for example in the decoherence regime for the gamma distribution discussed in Section \ref{decoherence}, or in the equilibrate state for the Lévy-Smirnoff distribution). 
In such a case the purity reaches its minimum value $\mathcal P_{\min}=\mathcal P [\hat\rho(0)]-\Delta \mathcal P_{\max}$ and
\beq\label{purmin}
\frac{1}{d}\leq \mathcal{P}_{\min}(\hat\rho_{\rm sr})=\mathcal{P}(\hat\rho_{\rm sr}=\hat\rho_{D})=\sum_{n}|c_n|^4=1-S_L\{|c_n|^2\},
\eeq
where $d$ is the effective dimension of the Hilbert space, or rather the number of terms in the expansion (\ref{psi}). The lower bound is attained only for equally-weighted superpositions (\ref{psi}) such that $|c_n|^2=1/d$. 
These superpositions are therefore the most affected pure states under stochastic resetting with the gamma (in particular the exponential) distribution in the decoherence regime, and also with the Lévy-Smirnoff distribution in the long-time regime. 
  
Finally, for the general equilibrate state given by (\ref{eqState}), its purity follows directly from \Eref{purityA} and reads
\beq
\mathcal{P}(\op{\rho}_{\rm sr,eq})=\sum_{n}[\rho_{nn}(0)]^2+{\mathsf L}^{2}\sum_{n\neq m}|\widetilde{\mathcal{T}}_{0}(i\omega_{nm})|^2|\rho_{nm}(0)|^2.
\eeq
The first term in the right-hand side corresponds to the purity of $\op{\rho}_{\rm D}$ and settles the minimum of $\mathcal{P}(\op{\rho}_{\rm sr,eq})$, attained when complete decoherence occurs ($\mathsf{L}=0$). The second term, identified with the contribution to $\mathcal{P}(\op{\rho}_{\rm sr,eq})$ due to stochastic resetting, can be bounded as
\begin{equation}
0\le{\mathsf L}^{2}\sum_{n\neq m}|\widetilde{\mathcal{T}}_{0}(i\omega_{nm})|^2|\rho_{nm}(0)|^2\le \sum_{n}\sigma^2_{n}[\op{\rho}(0)],    
\end{equation}
where $\sigma^2_{n}[\op{\rho}(0)]=\Mean{n}{\op{\rho}^{2}(0)}-\Mean{n}{\op{\rho}(0)}^{2}$ stands for the variance of $\op{\rho}(0)$ in the energy eigenstate $\ket{n}$. The reduction in the purity of the state due to the resetting goes in line with the idea that the renewal process can be interpreted as an effective non-unitary dynamics, and simulates the loss of information from the system to its surroundings $E$ (recall the discussion in the last paragraph of Section \ref{decoherence}), ascribable to the presence of correlations between them.


\paragraph{Fidelity.-} The `similarity' between the states $\op{\rho}_{\rm sr}$ and $\op{\rho}$ can be quantified by means of an appropriate fidelity measure $\mathcal{F}(\op{\rho}_{\rm sr},\op{\rho})$, which reaches its maximum value ($1$) whenever the states are indistinguishable, and its minimum ($0$) if the states are distinguishable.
Although the most widely known fidelity measure is the Uhlmann-Josza $\mathcal{F}_{\rm UJ}(\op{\varrho},\op{\varsigma})=\Bigl(\rm{Tr}\sqrt{\sqrt{\op{\varrho}}\,\op{\varsigma}\sqrt{\op{\varrho}}}\Bigr)^2$, it has computational disadvantages that can be avoided resorting to other fidelity measures \cite{LiangRPP2019}. Here we will focus on the fidelity $\mathcal{F}(\op{\varrho},\op{\varsigma})=\Tr\,(\op{\varrho} \,\op{\varsigma})/\max\{\mathcal P(\op{\varrho}),\mathcal P(\op{\varsigma})\}$, with $\mathcal P(\op{\varrho})$ the purity of the state $\hat \varrho$. Recalling that $\mathcal P (\op{\rho}_{\rm sr})\leq \mathcal P (\op{\rho})$, the fidelity between the states $\op{\rho}_{\rm sr}(t)$ and $\op{\rho}(t)$ is therefore
\begin{numparts}
\begin{eqnarray}
\mathcal{F}[\op{\rho}_{\rm sr}(t),\op{\rho}(t)]&=&\frac{\Tr\,[\op{\rho}_{\rm sr}(t)\,\op{\rho}(t)]}{\mathcal P [\hat \rho(0)]}=
\frac{1}{\mathcal P [\hat \rho(0)]}\sum_{n,m}\langle n|\hat \rho_{\rm sr}(t)|m\rangle \langle m|\hat \rho(t)|n\rangle
\nonumber\\
&=&\frac{1}{\mathcal P [\hat \rho(0)]}\sum_{n,m}|\rho_{nm}(0)|^2 I_{nm}(t)\label{fid2}\\
&=&1-\Delta \mathcal{F}(t),\label{fid3}
\end{eqnarray}
\end{numparts}
where 
\beq \label{DeltaF}
\Delta \mathcal{F}(t)=
\frac{1}{\mathcal P [\hat \rho(0)]}\sum_{n\neq m}|\rho_{nm}(0)|^2 (1-{\rm Re}\, I_{n m}(t))
\eeq
stands for the deviation of $\op{\rho}_{\rm sr}$ from the unitarily evolved state $\hat \rho$. 
As with the minimum purity, the minimum fidelity is attained whenever the reset state becomes an incoherent one (so $I_{nm}$ vanishes for all $n\neq m$), and reduces to
\beq
\fl\mathcal{F}_{\min}[\op{\rho}_{\rm sr},\op{\rho}]=\mathcal{F}[\op{\rho}_{\rm sr}=\op{\rho}_{\rm D},\op{\rho}]=\frac{1}{\mathcal P [\hat \rho(0)]}\sum_n[\rho_{nn}(0)]^2=\frac{1}{\mathcal P [\hat \rho(0)]}(1-S_L[\{\rho_{nn}\}]).
\eeq
Notice that if the initial state is pure, then $\mathcal{F}_{\min}$ coincides with $\mathcal{P}_{\min}$, given by Eq. (\ref{purmin}). 
Further, if the pure state (\ref{psi}) is the equally-weighted superposition ($|c_n|^2=1/d$), then $\mathcal{F}_{\min}=1/d$, and consequently as the number of accessible levels increases ($d\rightarrow \infty$) the fidelity between the incoherent state $\op{\rho}_{\rm D}$ and the unitarily evolved one tends to zero. This occurs, in particular, in the long-time regime when the equally-weighted superposition is subject to the resetting with the L\'evy-Smirnoff distribution.  

If we consider instead the fidelity between the reset state and the initial one, we find
\beq
\mathcal{F}[\op{\rho}_{\rm sr}(t),\op{\rho}(0)]=\frac{\Tr\,[\op{\rho}_{\rm sr}(t)\,\op{\rho}(0)]}{\mathcal P [\hat \rho(0)]}=
\frac{1}{\mathcal P [\hat \rho(0)]}\sum_{n,m}|\rho_{nm}(0)|^2 \mathcal{I}_{nm}(t).
\eeq
Naturally this expression has the same structure as (\ref{fid2}) but now involving the factor $\mathcal{I}_{nm}(t)$, which bears the effects of both the resetting \emph{and} the unitary evolution. The deviation of $\op{\rho}_{\rm sr}$ from the initial state $\hat \rho(0)$ is thus obtained from Eq. (\ref{DeltaF}) replacing $I_{nm}(t)$ by $\mathcal{I}_{nm}(t)$. In particular, in the Zeno regime of the exponential resetting example introduced in the previous section, direct calculation shows that ${\rm Re}\,\mathcal{I}_{nm}(t)$ tends to unity as $\alpha$ increases, consequently the corresponding deviation $\Delta \mathcal{F}$ tends to zero, and $\mathcal{F}\bigl[\op{\rho}_{\rm sr}(t),\hat\rho(0)]\rightarrow 1$, as expected. 

In the (long-time) equilibration regime, when $\op{\rho}_{\rm sr}(t)$ has relaxed to $\op{\rho}_{\rm sr,eq}$, we resort to Eq. (\ref{eqState}) and find that 
\begin{eqnarray}\label{fieq}
\fl\mathcal{F}[\op{\rho}_{\rm sr,eq},\hat \rho(t)]&=&\frac{1}{\mathcal P [\hat \rho(0)]}\Bigl[\sum_{n}[\rho_{nn}(0)]^2+{\mathsf L}\sum_{n\neq m}\widetilde{\mathcal{T}}_{0}(i\omega_{nm})\vert\rho_{nm}(0)\vert^{2} e^{i\omega_{nm}t}\Bigr],
\end{eqnarray}
so the fidelity oscillates around its minimum value $\mathcal{F}_{\min}$.


\section{Evolution of a qubit under stochastic resetting}

Systems with an effective 2-dimensional Hilbert space, or qubits, deserve particular attention mainly because they are paradigmatic systems in quantum information protocols, and admit a simple mathematical description. 
Throughout this section we thus analyse the evolution of qubits under the stochastic resetting process. Firstly, we focus on pure states, specifically on states of the form (\ref{psi}) with only two components. Afterwards, in Subsection \ref{Bloch}, we drop out this restriction and consider general (mixed) qubit states undergoing generic dynamical maps as well as unitary evolutions.

Equations (\ref{DeltaP}) and (\ref{DeltaF}) become particularly simple when applied to superpositions of only two energy eigenstates, say the $i$-th and the $j$-th ($i,j\in\{1,\dots,N\}$ with $N$ the number of non-degenerate eigenstates of $\hat H$), and read 
\begin{numparts}
\begin{eqnarray}\label{DeltaPqubit}
\Delta \mathcal{P}_{ij}(t)&=&2|c_i|^2 (1-|c_i|^2)[1-|I_{ij}(t)|^2]\leq \frac{1}{2}[1-|I_{ij}(t)|^2],\\
\Delta \mathcal{F}_{ij}(t)&=& 2|c_i|^2 (1-|c_i|^2)[1-{\rm Re}\,I_{ij}(t)]\leq \frac{1}{2}[1-{\rm Re}\,I_{ij}(t)]. \label{DeltaFqubit}
\end{eqnarray}
\end{numparts}
The inequalities saturate for $|c_i|^2=1/2$, which means that, for fixed resetting distributions and energy spectrum, the (pure) qubit states that are more affected by stochastic resetting (which are more deviated from the unitarily evolved state, and become more mixed) are the equally-weighted superpositions of energy eigenstates
\beq\label{psiqubit}
\ket{\psi{(0)}}=\frac{1}{\sqrt2}(\ket{i}+e^{i\varphi}\ket{j})\equiv\ket{\psi^{(ij)}}_{\rm q}.
\eeq
Unbalanced superpositions tend to be more robust under the resetting as the unbalance is increased, until the most asymmetric superposition, with $c_n=\delta_{nk}$, is completely unaffected since it corresponds to a stationary state.

The states (\ref{psiqubit}) have also the distinctive property of evolving, under arbitrary $\hat U=e^{-i\hat H t/\hbar}$, into an orthogonal state in the minimum possible time allowed by the \emph{quantum speed limit} \cite{LevitinPRL2009,DeffnerJPhysA2017}. The corresponding \emph{orthogonality time} $\tau_{ij}$, needed to transform the state $|\psi^{(ij)}\rangle_{\rm q}$ into and orthogonal one $|\psi^{(ij)}_{\perp}\rangle_{\rm q}$, coincides thus with the Mandelstam-Tamm \cite{Mandelstam1991} and the Margolus-Levitin \cite{MargolusPhysicaD1998} bounds, and is given by
\beq\label{qsl}
\tau_{ij}=\frac{\pi}{|\omega_{ij}|}.
\eeq
Clearly, states determined by different pairs $i,j$ evolve with different speed, which increases with the energy dispersion of the state. Accordingly, if the Hamiltonian eigenstates are ordered as $E_1<\dots<E_N$, the superposition corresponding to $i=1,j=N$ reaches its orthogonal state faster than any other one, and the superposition involving the closest adjacent states is the slowest \cite{ValdesJPhysA2020}. 

Now, for all states $|\psi^{(ij)}\rangle_{\rm q}$ we haven seen that $\Delta \mathcal{F}_{ij}(t)$ and $\Delta \mathcal{P}_{ij}(t)$ become maximal, but such maximum values depend on $\omega_{ij}$, or rather on the corresponding orthogonality time. 
It is therefore natural to inquire whether the speed of evolution towards orthogonality provides (or not) more robustness on the fidelity and the purity under the effects of stochastic resetting. With this aim we first notice that for sufficiently small times such that $|\omega_{ij}|t\ll 1$ we can write
\begin{numparts}
\begin{eqnarray}
{\rm Re}\,I_{ij}(t)&\approx& 1-\omega^2_{ij}\,\mathsf R(t)\quad{\rm with}\quad \mathsf R(t)=\frac{1}{2}\int_0^t ds\,\mathcal{T}(s)\mathcal{T}_0(t-s)s^2,\\
{\rm Im}\,I_{ij}(t)&\approx& \omega_{ij}\,\mathsf I(t)\quad{\rm with}\quad \mathsf I(t)=\int_0^t ds\,\mathcal{T}(s)\mathcal{T}_0(t-s)s,
\end{eqnarray}
\end{numparts}
and consequently, to lowest order in $|\omega_{ij}|t$ we get
\beq\label{ReIm}
1-{\rm Re}\,I_{ij}(t)\approx\omega^2_{ij}\,\mathsf R(t),\quad 1-|I_{ij}(t)|^2\approx \omega^2_{ij}\,[2\mathsf R(t)-\mathsf I^2(t)].
\eeq
The Cauchy-Schwarz inequality gives $2\mathsf R(t)-\mathsf I^2(t) \geq 0$, whence Eqs. (\ref{ReIm}) together with (\ref{DeltaPqubit}) and (\ref{DeltaFqubit}) imply that for two states of the form (\ref{psiqubit}), namely $|\psi^{(ij)}\rangle_{\rm q}$ and $|\psi^{(kl)}\rangle_{\rm q}$ with $|\omega_{ij}|<|\omega_{kl}|$, it holds that $\Delta\mathcal{P}_{ij}<\Delta\mathcal{P}_{kl}$, and $\Delta\mathcal{F}_{ij}<\Delta\mathcal{F}_{kl}$. Therefore, for sufficiently small times $t$ such that $t\ll \tau_{kl}/\pi$ (so neither of the qubits has reached an orthogonal state) the fidelity and the purity are more affected for the faster state $|\psi^{(kl)}\rangle_{\rm q}$. In particular, from among all states of the form (\ref{psiqubit}) the one that is more drastically affected if the fastest one, $|\psi^{(1N)}\rangle_{\rm q}$.    

\subsection{Mean orthogonality time}

The effects of stochastic resetting on the otherwise unitary evolution of the state $|\psi^{(ij)}\rangle_{\rm q}$ are clearly to hinder the transition towards the orthogonal one $|\psi^{(ij)}_{\perp}\rangle_{\rm q}$. In such a situation the orthogonality time becomes a stochastic variable $\tau_{\rm sr}$, and to extract information regarding the actual delay in arriving at an orthogonal state we compute the 
\emph{mean orthogonality time} under resetting,
 $\langle\tau_{\rm sr}\rangle$, for arbitrary stochastic resetting dynamics, that is, for arbitrary distribution of times between resetting events $\mathcal{T}_{1}(t)$.

\begin{figure}\centering
    \includegraphics[width=0.395\textwidth,trim=100 0 77 0,clip=true]{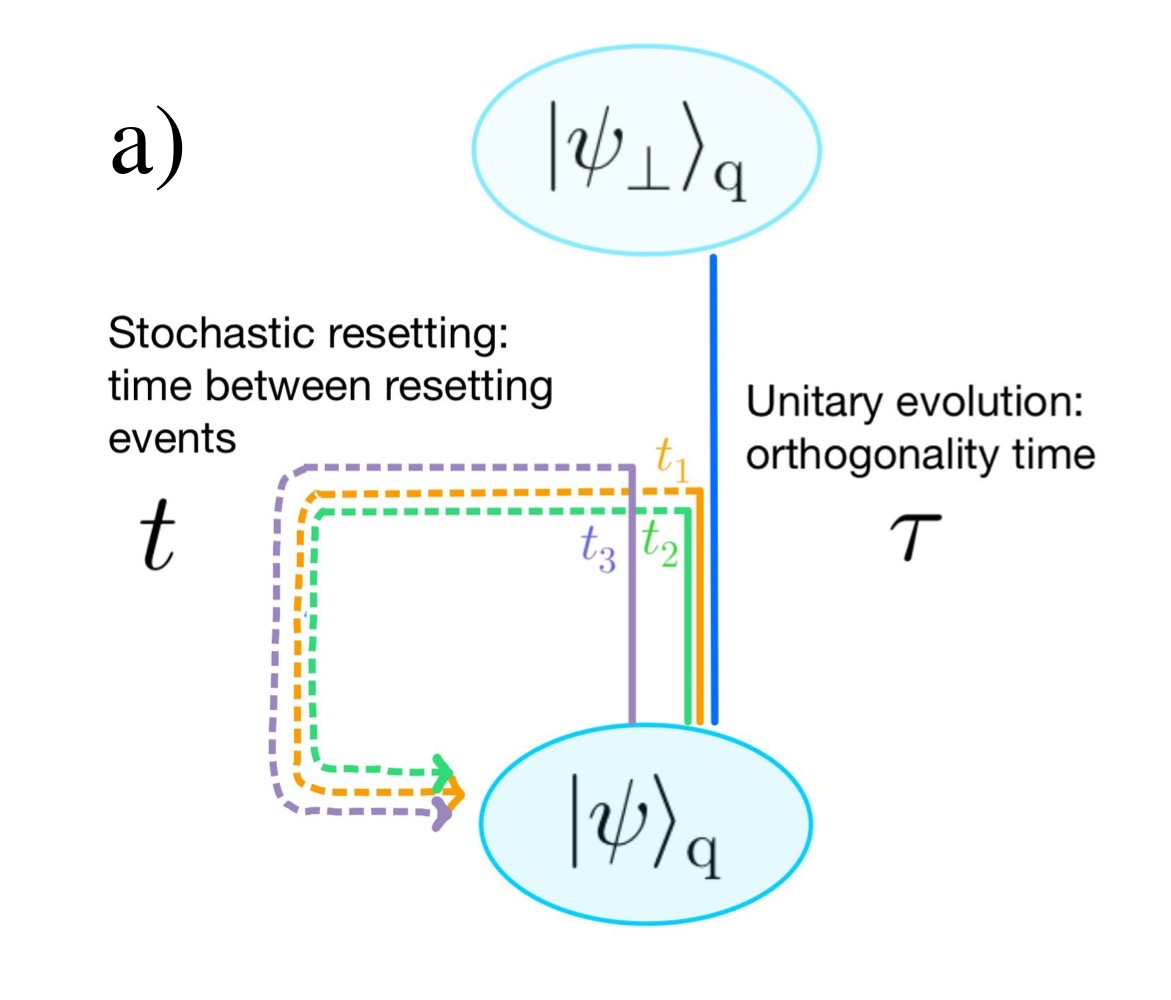}    \includegraphics[width=0.595\textwidth,trim=0 100 0 0]{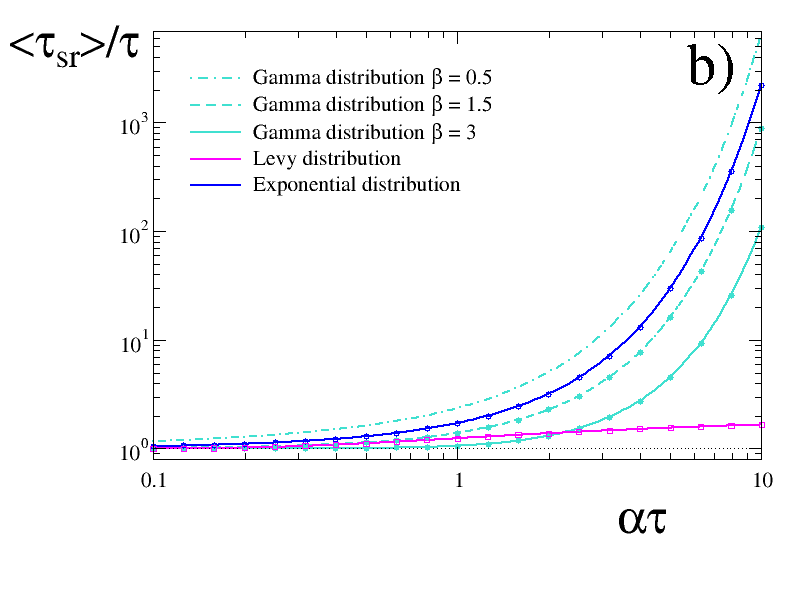}
    \caption{a) In absence of stochastic resetting, the initial state $\ket{\psi}_{\rm q}$ evolves unitarily until it reaches the orthogonal state $\ket{\psi_{\perp}}_{\rm q}$ for the first time at $\tau=\pi/\omega$. Stochastic resetting can delay the evolution towards $\ket{\psi_{\perp}}_{\rm q}$ since it restarts the state evolution repeatedly whenever $t$, drawn from $\mathcal{T}_{1}(t)$, is smaller than $\tau$. The orthogonal state is reached provided the time $t$ between consecutive events satisfies $t\ge\tau$. b) The mean orthogonality time $\langle\tau_{\rm sr}\rangle$ in units of $\tau$ as function of $\alpha\tau$ is shown from numerical calculations (symbols) and evaluation of the analytical expressions (lines) given in Table \ref{tabla} for: the exponential distribution (blue); the gamma distribution (turquoise); and the Lévy-Smirnoff distribution (magenta). Here $\alpha$ defines the mean time between resetting events (exponential and gamma distributions) or defines the maximum of the distribution (Lévy-Smirnoff distribution).}
    \label{fig:MeanOrthogonalityTime}
\end{figure}

The calculation of $\langle\tau_{\rm sr}\rangle$ belongs to the class of \emph{mean first passage time} processes discussed in Refs. \cite{ReuveniPRL2016,PalPRL2017}. Consider the time evolution of an initial state of the form \eref{psiqubit}, which for simplicity we denote as  $\ket{\psi}_{\rm q}$. In the absence of resetting events it reaches the orthogonal state $\ket{\psi_{\perp}}_{\rm q}$ for the first time at time $\tau$ (blue line in Fig. \ref{fig:MeanOrthogonalityTime}.a), however its attainment is delayed when the evolution is interrupted and restarted at random times $t_i<\tau$ (color lines in Figure \ref{fig:MeanOrthogonalityTime}.a). Thus, if the system reaches the orthogonal state prior to the first resetting event the orthogonality time is simply $\tau$, otherwise, the evolution will be repeatedly restarted from $\ket{\psi}_{\rm q}$ 
 until the time between consecutive resets is greater than $\tau$. Let us denote with $\tau_{\rm sr}$ the (random) time it takes for the system to attain the state $\ket{\psi_{\perp}}$ for the first time, and with $t$ the time between consecutive resetting events, we are thus led to 
the recurrence relation \cite{ReuveniPRL2016}
\begin{equation}\label{TauSR}
\tau_{\rm sr}=
\cases{\tau & if $t\ge\tau$ \\ 
t+\tau^{\prime}_{\rm sr} & if $t<\tau$},
\end{equation}
where $\tau^{\prime}_{\rm sr}$ is a stochastic variable with the same statistical properties of 
$\tau_{\rm sr}$, thus also satisfying 
\Eref{TauSR}.

The latter can be rewritten as $\tau_{\rm sr}={\rm min}(\tau,t)+\tau^{\prime}_{\rm sr}\, \theta(\tau-t)$ \cite{PalPRL2017}, where $\theta(z)$ denotes the Heaviside step function, being 1 if $z>0$ and 0 if $z\le0$. From this the probability density of $\tau_{\rm sr}$ is $P(\tau_{\rm sr})=\int_{0}^{\infty}dt\, \mathcal{T}_1(t)\delta\bigl[\tau_{\rm sr}-{\rm min}(\tau,t)-\tau^{\prime}_{\rm sr}\, \theta(\tau-t)\bigr]$, and the mean orthogonality time thus satisfies
\beq
\mean{\tau_{\rm sr}}=\int_0^\infty dt\,\mathcal{T}_{1}(t)\,{\rm min}(\tau,t)+\mean{\tau_{\rm sr}}\int_{0}^{\infty}dt\,\mathcal{T}_{1}(t)\, \theta(\tau-t),
\eeq
from which we get
\begin{equation}\label{AnalyticalMOT}
\langle\tau_{\rm sr}\rangle=\frac{\displaystyle\int_{0}^{\infty}dt\, \mathcal{T}_{1}(t)\, {\rm min}(\tau,t)}{1-\displaystyle\int_{0}^{\infty}dt\, \mathcal{T}_{1}(t)\theta(\tau-t)}
=\tau+\frac{\displaystyle\int_{0}^{\tau}dt\, \mathcal{T}_{1}(t)\, t}{\displaystyle\int_{\tau}^{\infty}dt\, \mathcal{T}_{1}(t)}.
\end{equation}
This last expression allows to quantify the delay in the orthogonality time and elucidates the influence caused by resetting through $\mathcal{T}_1(t)$. In Figure \ref{fig:MeanOrthogonalityTime}.b we show the mean orthogonality time (in units of $\tau$) as function of $\alpha$ (in units of $\tau^{-1}$), which characterises the $\mathcal{T}_1$'s discussed in Section \ref{subsect:ParticularProcesses}: Lines depict the analytical expressions obtained from \Eref{AnalyticalMOT} and explicitly written in Table \ref{tabla}, while symbols depict numerical calculations\footnote{Numerical simulations were carried out by considering a sample of $5\times 10^{5}$ successful events for which the system reaches the orthogonal state, for each $\mathcal{T}_1$ considered.}. For $\mathcal{T}_1$'s with $\alpha\lesssim\tau$, the mean orthogonality times stays of the order of $\tau$. In contrast, for $\alpha\gtrsim\tau$, the gamma distribution gives the larger delay effects on $\mean{\tau_{\rm sr}}$ as the the parameter $\beta$ decreases, while for the Lévy-Smirnoff distribution $\mean{\tau_{\rm sr}}$ slowly increases with $\alpha$ to saturate at the value $\alpha=2\tau$, thus driving the system to $\ket{\psi_{\perp}}_{\rm q}$ in the least mean time.
\Table{\label{tabla} Analytical expressions for the mean orthogonality time, $\langle\tau_{\rm sr}\rangle$, for the three probability densities $\mathcal{T}_{1}(t)$ considered in the text.}
\br
\centre{2}{$\mathcal{T}_{1}(t)$}&\centre{2}{$\langle\tau_{\rm sr}\rangle$}\\
\mr
\centre{2}{$\alpha e^{-\alpha t}$}&\centre{2}{$\displaystyle\frac{1}{\alpha}(e^{\alpha\tau}-1)$}\\
\centre{2}{$\displaystyle\frac{\alpha}{\Gamma(\beta)}t^{\beta-1} e^{-\alpha t}$}&\centre{2}{$\displaystyle\tau+\frac{1}{\alpha}\frac{\gamma(\beta+1,\alpha\tau)}{\Gamma(\beta,\alpha\tau)}^{\,\,\rm a}$}\\
\centre{2}{$\displaystyle \frac{e^{-\frac{1}{2\alpha t}}}{\sqrt{2\pi\alpha t^{3}}}$}&\centre{2}{\qquad$\displaystyle\tau+\frac{1}{\alpha}+\frac{1}{\alpha\, {\rm erf}\Bigl(\frac{1}{\sqrt{2\alpha\tau}}\Bigr)}\biggl[\frac{\sqrt{2\alpha\tau}}{\sqrt{\pi}}e^{-\frac{1}{2\alpha\tau}}-1\biggr]$}$^{\,\, \rm b}$\\
\br
\end{tabular}
\item[] $^{\rm a}$ $\gamma(x,\sigma)=\int_{0}^{x}dz\, z^{\sigma-1}e^{-z}$ and $\Gamma(x,\sigma)=\int_{x}^{\infty}dz\, z^{\sigma-1}e^{-z}$ denote the lower and upper incomplete gamma function, respectively.
\item[] $^{\rm b}$ ${\rm erf}(x)=\frac{2}{\sqrt\pi}\int_0^x dz\, e^{-z^2}$ is the error function.
\end{indented}
\end{table}




\subsection{Bloch vector dynamics}\label{Bloch}

In this section we explicitly examine the effects of stochastic resetting on the evolution of a single-qubit generic state $\hat \varrho$, that may be either pure or mixed. The complete description of the state can be carried out resorting to the Bloch sphere, 
which consists of the vectors defined as $\boldsymbol{r}_{\varrho}=\rm{Tr}\,(\op{\varrho}\,\op{\boldsymbol{\sigma}})$ (with $\op{\boldsymbol{\sigma}}=(\hat \sigma_x,\hat \sigma_y,\hat \sigma_z)$, $\{\hat \sigma_i\}$ being the Pauli matrices), with $\vert\boldsymbol{r}_{\varrho}\vert\le1$. In terms of the \emph{Bloch vector} $\boldsymbol{r}_{\varrho}$ the state $\op{\varrho}$ is  
\begin{equation}\label{rhoqubit}
\op{\varrho}=\frac{1}{2}\bigl(\mathbb{I}+\boldsymbol{r}_{\varrho}\cdot\op{\boldsymbol{\sigma}}\bigr),
\end{equation}
and its purity can be written as 
\beq
{\rm Tr}\,\op{\varrho}^2=\frac{1}{2}\bigl(1+\vert\boldsymbol{r}_{\varrho}\vert^2\bigr),
\eeq
so unit vectors correspond to pure states whereas  $\vert\boldsymbol{r}_{\varrho}\vert<1$ holds for mixed ones. Inspection of the Bloch vector thus provides all the information of the qubit state. 

Let
\begin{equation}
\hat \rho(t)=\frac{1}{2}\bigl[\mathbb{I}+\boldsymbol{r}(t)\cdot\op{\boldsymbol{\sigma}}\bigr]
\end{equation}
denote the non-reset state evolving under the dynamic map $\mathcal{E}$. By use of (\ref{DensityOP}) 
the correspondingly state under stochastic resetting is \begin{equation}\label{densityBloch}
\op{\rho}_{\rm{sr}}(t)=\frac{1}{2}\bigl[\mathbb{I}+\boldsymbol{r}_{\rm{sr}}(t)\cdot\op{\boldsymbol{\sigma}}\bigr]
\end{equation}
with
\beq \label{blochvectorSR}
\boldsymbol{r}_{\rm{sr}}(t)=\mathcal{T}_{0}(t)\boldsymbol{r}(t)+\int_{0}^{t}ds\, \mathcal{T}(s)\mathcal{T}_{0}(t-s)\boldsymbol{r}(t-s),
\eeq
whence $|\boldsymbol{r}_{\rm{sr}}(t)|\leq |\boldsymbol{r}(t)|$,
and the purity decreases as a result of resetting for \emph{all} dynamical maps $\mathcal{E}$.

We now 
center our analysis to the case for which $\mathcal{E}$ is represented by a unitary Hamiltonian evolution driven by
\begin{equation}
\op{H}=\frac{1}{2}(e_{0}\mathbb{I}+\boldsymbol{e}\cdot\op{\boldsymbol{\sigma}}),
\end{equation}
where $e_0=\rm{Tr}\,\op{H}$ and  $\boldsymbol{e}=\rm{Tr}\,(\op{H} \op{\boldsymbol{\sigma}})$. The eigenvalues of $\op{H}$ are $E_{\pm}=(e_0\pm e)/2$, with $e=\vert\boldsymbol{e}\vert$. 
Without loss of generality we set $e_0=0$ and introduce $\omega=(E_+ - E_-)/\hbar=e/\hbar$. For an initial state $\op{\rho}(0)$ with Bloch vector $\boldsymbol{r}_0=(x_0,y_0,z_0)$, the unitary evolution \eref{RhoUnitary} is accomplished by the SU(2) rotation $\hat U=\op{R}(\boldsymbol{n},\theta)=\exp\{-i\theta\boldsymbol{n}\cdot\op{\boldsymbol{\sigma}}/2\}$, around the direction $\boldsymbol{n}=\boldsymbol{e}/e$ by the angle $\theta=\omega t$, while the evolution of the corresponding Bloch vector is accomplished by a proper rotation $\mathbb{R}(\boldsymbol{n},\theta)$ in SO(3) around the same direction $\boldsymbol{n}$ and by the same angle $\theta$,
thus $\boldsymbol{r}(t)=\mathbb{R}(\boldsymbol{n},\omega t)\, \boldsymbol{r}_{0}$ describes a vector precessing around $\boldsymbol{n}$.
The stochastic resetting interrupts the precession dynamics of $\boldsymbol{r}(t)$, restarting the evolution from $\boldsymbol{r}_{0}$ immediately after each resetting event, and the evolution of the reset Bloch vector can be written, with the aid of \Eref{blochvectorSR}, as
\begin{equation}\label{blochvectorSR2}
\boldsymbol{r}_{\rm sr}(t)=\mathcal{R}(\boldsymbol{e}/e,\omega t)\, \boldsymbol{r}_0 
\end{equation}
where the transformation $\mathcal{R}(\boldsymbol{e}/e,\omega t)$ is non-orthogonal and given by
\begin{equation}
\mathcal{R}(\boldsymbol{e}/e,\omega t)= \mathcal{T}_{0}(t)\mathbb{R}(\boldsymbol{e}/e,\omega t)+\int_{0}^{t}ds\, \mathcal{T}(s)\mathcal{T}_{0}(t-s)\mathbb{R}[\boldsymbol{e}/e,\omega(t-s)].
\end{equation}





Finally, explicit analytical expressions for the components of the Bloch vector $\boldsymbol{r}_{\rm sr,eq}={\rm Tr}(\op{\rho}_{\rm sr,eq}\,\op{\boldsymbol{\sigma}})$ associated to the equilibrate state \eref{eqState} can be computed for $\mathcal{T}_{1}$ given by the gamma distribution \eref{GammaDist} when, without loss of generality, we choose $\op{H}=\frac{e}{2}\op{\sigma}_{z}$. 
We have for the initial qubit $\op{\rho}(0)$, characterised by the Bloch vector
$\boldsymbol{r}_0=
(x_{0},y_{0},z_{0})$, that:
\begin{numparts}
\begin{eqnarray}
\fl x_{\rm sr,eq}&=& x_{0}\frac{1}{\beta}\frac{\alpha}{\omega}\frac{\alpha^{\beta}}{(\alpha^{2}+\omega^{2})^{\beta/2}}\sin(\Theta\beta)+y_0\frac{1}{\beta}\frac{\alpha}{\omega}\Bigl[\frac{\alpha^{\beta}}{(\alpha^{2}+\omega^{2})^{\beta/2}}\cos(\Theta\beta)-1\Bigr],\\
\fl y_{\rm sr,eq}&=& y_{0}\frac{1}{\beta}\frac{\alpha}{\omega}\frac{\alpha^{\beta}}{(\alpha^{2}+\omega^{2})^{\beta/2}}\sin(\Theta\beta)+x_0\frac{1}{\beta}\frac{\alpha}{\omega}\Bigl[1-\frac{\alpha^{\beta}}{(\alpha^{2}+\omega^{2})^{\beta/2}}\cos(\Theta\beta)\Bigr],\\
\fl z_{\rm sr,eq}&=&z_{0},
\end{eqnarray}
\end{numparts}
where $\Theta=\arctan(\omega/\alpha)$.

\subsection{An example: Two-level atom interacting with a quantized electromagnetic field}
We will now study the effects of stochastic resetting on a  two-level atom system that interacts with a single mode of the quantized electromagnetic field, described by the paradigmatic   Jaynes-Cummings Hamiltonian \cite{AolitaRPP2015}
\beq\label{Hatomo}
\hat H_{\rm{JC}}=\hbar \omega\Big(\hat a^{\dagger}\hat a+\frac{1}{2}\Big)-\frac{1}{2}\hbar \omega \hat \sigma_z-i\frac{1}{2}\hbar \Omega\,(\hat \sigma^{+}\hat a-\hat \sigma^{-}\hat a^{\dagger}).
\eeq
Here $\hat a^{\dagger}$ and $\hat a$ stands, respectively, for the creation and annihilation operators of the electromagnetic field, $\hat \sigma^{+}$ and $\hat \sigma^{+}$ for the raising and lowering operators of the atom, $\omega$ is the frequency of the electromagnetic mode, which coincides with the atomic transition frequency between the excited and the ground atomic states, and $\Omega$ is the real coupling constant. The first two terms in (\ref{Hatomo}) correspond to the free Hamiltonians of the field and atom, respectively, whereas the last term (the interaction Hamiltonian $\op{H}_{\rm int}$) allows for the excitation (de-excitation) of the atom and the concomitant absortion (emission) of a photon.

We assume the atom$+$field initial state to be
\beq \label{psit0}
\ket{\psi(0)}=(a\ket{\downarrow}+b\ket{\uparrow})\ket{0},
\eeq
with $\ket{\downarrow}$ and $\ket{\uparrow}$ the ground and excited states of the atom, which are eigenvectors of $\hat \sigma_z$ with positive and negative eigenvalues, respectively. The vector $\ket{0}$ denotes the vacuum state of the field, while $\ket{1}$ will denote the excited state of the field with one photon. In the interaction picture the initial state evolves into
\beq\label{psievolved}
\ket{\psi(t)}=a\ket{\downarrow}\ket{0}+b\sqrt{1-p(t)}\ket{\uparrow}\ket{0}+b\sqrt{p(t)}\ket{\downarrow}\ket{1},
\eeq
where $p(t)=\sin^2(\Omega t/2)$.

We now focus on the (reduced) state of the two-level atom, defined as
\beq
\hat\rho^{A}(t)=\Tr _{F}\,\ket{\psi(t)}\bra{\psi(t)},
\eeq
where $\ket{\psi(t)}$ is given by (\ref{psievolved}), and ${\rm Tr}_{F}$ denotes the trace over the degrees of freedom of the field. Direct calculation gives
\begin{eqnarray}\label{rhoatom}
\hat \rho^{A}(p)&=&\Bigl(|a|^2+p(t)|b|^2\Bigr)\ket{\downarrow}\bra{\downarrow}+|b|^2\Bigl(1-p(t)\Bigr)\ket{\uparrow}\bra{\uparrow}\nonumber\\
&&+ab^*\sqrt{1-p(t)}\ket{\downarrow}\bra{\uparrow}+
a^*b\sqrt{1-p(t)}\ket{\uparrow}\bra{\downarrow}.
\end{eqnarray}
Although the evolution of the composite atom$+$field is unitary, the evolution of the atom alone is not; instead it is given by the dynamical map $
\hat \rho^{A}(p)=\mathcal{E}(p,0)\hat \rho^{A}(0)=\sum_{i=0,1}\hat K_i(p) \hat \rho^{A}(0)\hat K^{\dagger}_i(p),$
where $\{\hat K_i\}$ stands for the set of Kraus operators associated to the so-called \emph{amplitude damping} channel \cite{nielsen00,SallesPRA2008}. Assuming that $a,b\in \mathbb{R}$, and writing the operator (\ref{rhoatom}) in the form (\ref{rhoqubit}), we identify the corresponding Bloch vector as

\beq\label{rhoAunit}
\boldsymbol{r}^A(p)=\Bigl(2ab\sqrt{(1-p)},0,1-2b^2(1-p)\Bigr),
\eeq
so under the resetting the atom's state is $\hat \rho ^A_{\rm{sr}}(p)=\frac{1}{2}\bigl[\mathbb{I}+\boldsymbol{r}^{A}_{\rm{sr}}(p)\cdot\op{\boldsymbol{\sigma}}\bigr]$ with $\boldsymbol{r}^A_{\rm{sr}}(p)$ related to $\boldsymbol{r}^A(p)$ as indicated in Eq.   
(\ref{blochvectorSR}). 
\begin{figure}[!]
	\vspace{0.8cm}
	\begin{center}
\includegraphics[width=0.3\textwidth]{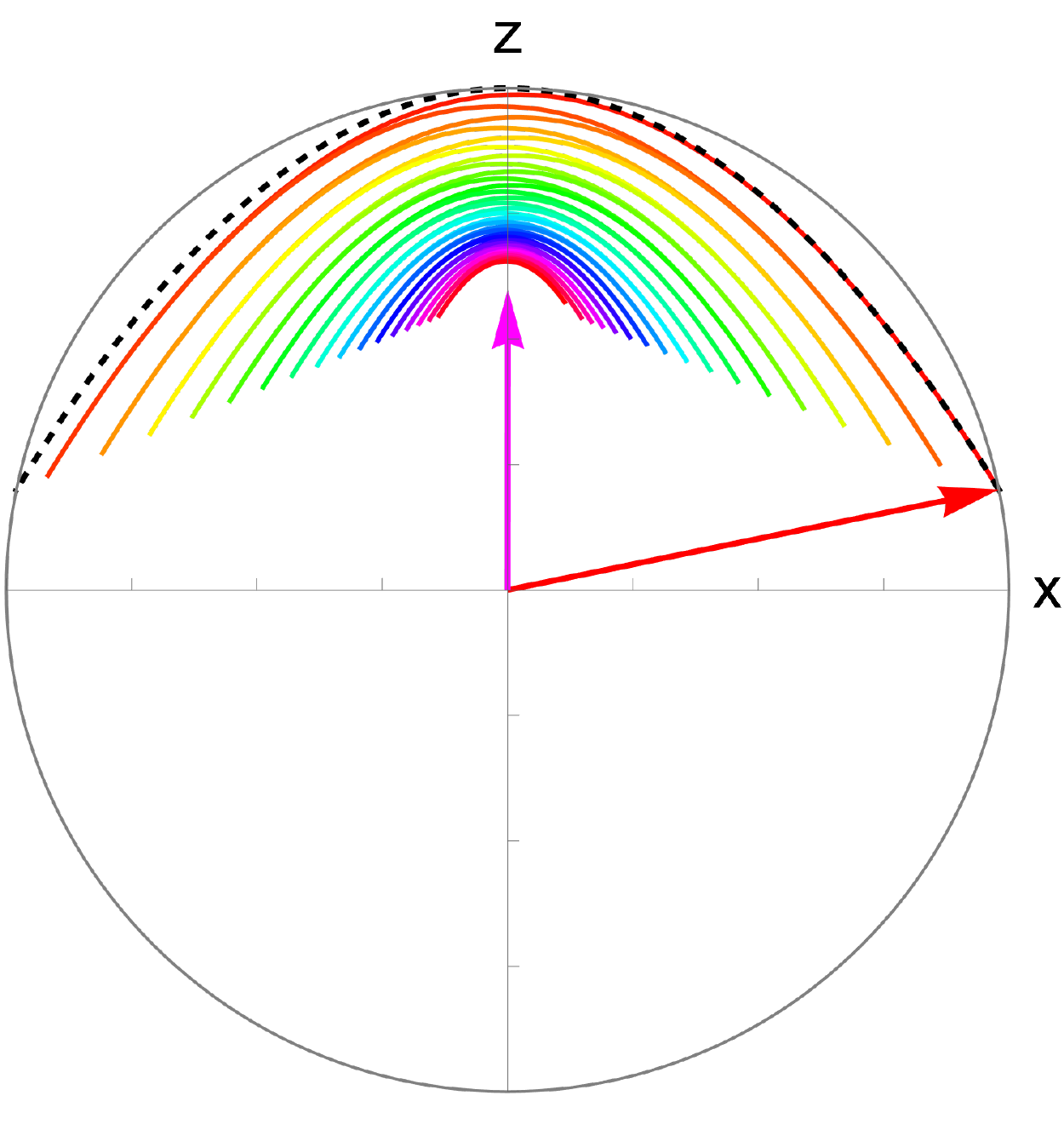}\quad\quad
\includegraphics[width=0.3\textwidth]{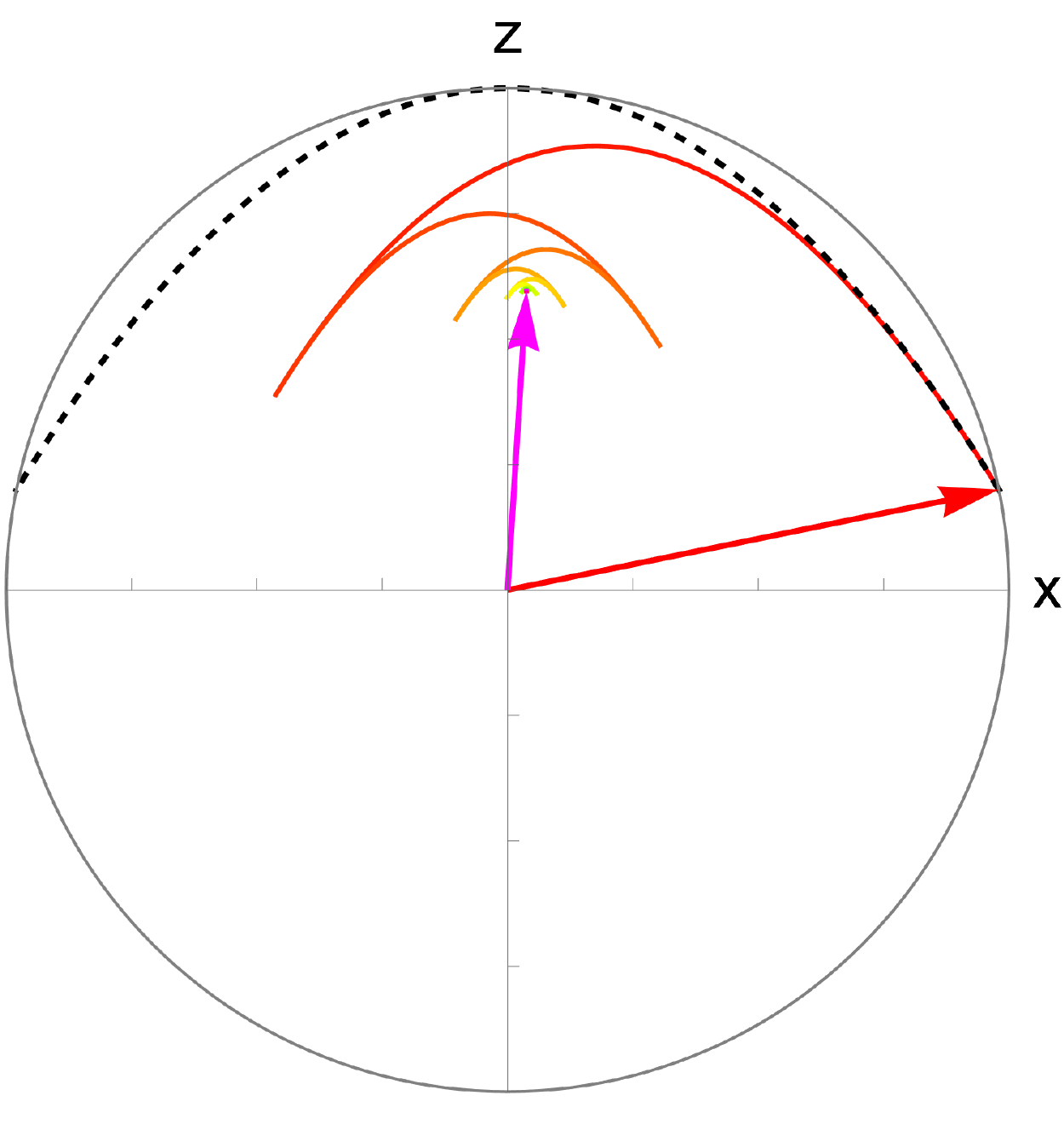}\\
\includegraphics[width=0.3\textwidth]{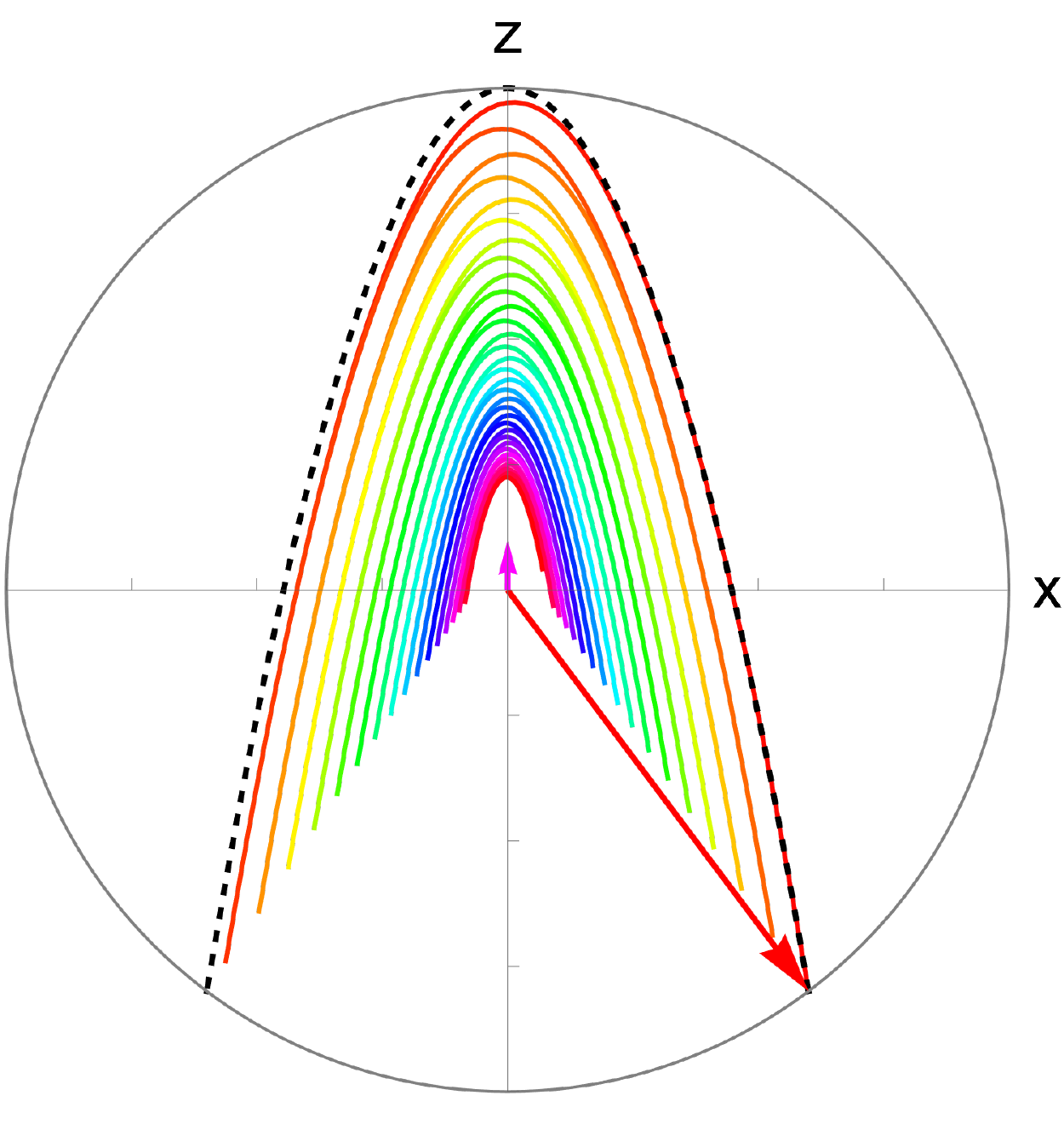}\quad\quad
\includegraphics[width=0.3\textwidth]{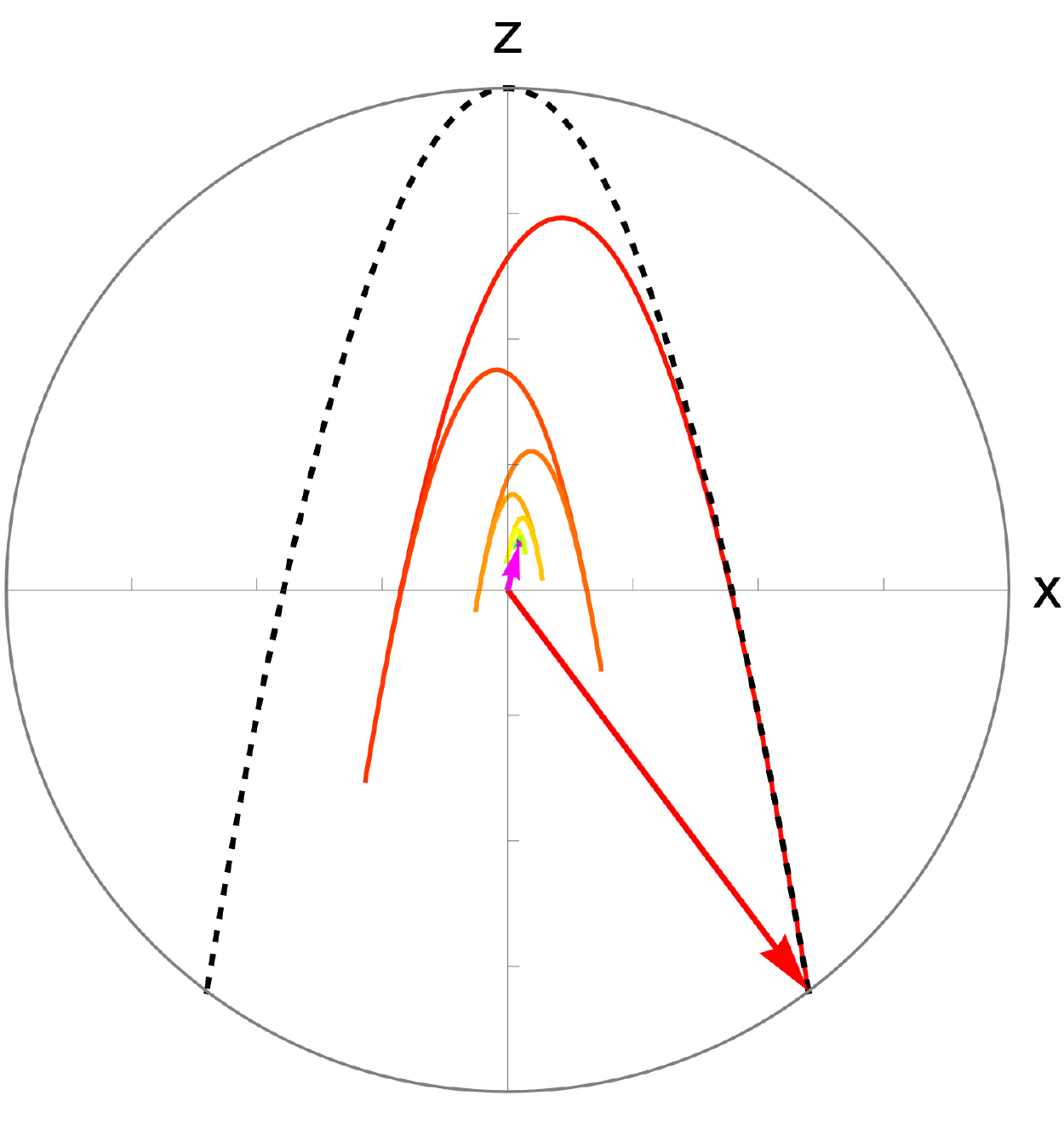}
\caption[]{\label{rhoA}Evolution of the Bloch vector $\boldsymbol{r}^A(t)=(2ab\cos{\Omega t/2},0,1-2b^2\cos^2{\Omega t/2})$ under stochastic resetting for: $a=\sqrt{3/5}$,  $b=\sqrt{2/5}$ (upper panel); $a=\sqrt{1/10}$,  $b=\sqrt{9/10}$ (bottom panel). The resetting corresponds to the exponential distribution with $\alpha=1/100$ (left panel) and $\alpha=1/10$ (right panel) in units of $\Omega$, and time runs from $t=0$ to $t=200$ (units of $\Omega^{-1}$), in color scale ranging from red to magenta. The initial ($t=0$) and the final ($t\rightarrow\infty$) Bloch vectors are indicated by a red and a magenta arrow, respectively. The black-dashed line depicts the periodic trajectory of the Bloch vector in absence of resetting ($\alpha=0$).
} 
	\end{center}
\end{figure}

The evolution of $\boldsymbol{r}^A_{\rm{sr}}(t)$ for the exponential distribution \eref{ExpDist} of resetting is shown in Fig. \ref{rhoA} for two different initial states (upper and lower panels) and two values of $\alpha$ (right and left panels).
As follows from Eqs. (\ref{rhoAunit}) and (\ref{blochvectorSR}), $\boldsymbol{r}^A_{\rm{sr}}(t)$ is restricted to the plane $x$-$z$, which is the one depicted in the figure. Each 
graph shows the evolution of the tip of the vector $\boldsymbol{r}^A_{\rm{sr}}(t)$, and time flow is represented by the color scale. The initial ($t=0$) and the final ($t\rightarrow \infty$) Bloch vectors are represented, respectively, by red and magenta arrows. For reference, the periodic trajectory (traversed back and forth) of the vector $\boldsymbol{r}^A(t)$ in absence of resetting is depicted with the black-dashed line. Notice that this trajectory corresponds to Bloch vectors that vary their magnitude as they evolve, evincing the fact that the evolution of $\op{\rho}^A(t)$ is a non-unitary one. 

Figure \ref{rhoA} verifies that an effect of the exponential resetting is to continuously reduce the magnitude of the Bloch vector ---and with it the purity of the state---, until a steady vector (state) is attained (notice that such limiting state is not the state $\op\rho_{\rm sr,eq}$ introduced in Sec. \ref{equilibrate_state}, since the latter assumed density matrices that evolve unitarily in the absence of resetting, whereas in this example, as already stated, the reduced state evolves in a non-unitary fashion). Comparison of the upper and lower panels shows that the magnitude of  $\boldsymbol{r}^A_{\rm sr}(t)$  differs more drastically from that of $\boldsymbol{r}^A(0)$ 
as the initial population $b^2$ of the atom's excited state $\ket{\uparrow}$ increases (see Eq. (\ref{psit0})).  
Notice also that the value of the resetting rate $\alpha$ affects not only the speed at which $\boldsymbol{r}^A_{\rm sr}(t)$ approximates its limiting value, but also such precise value.

\section{\label{section4}Final comments and concluding remarks}

We have studied the effects of stochastic resetting on the dynamics of closed quantum systems through a renewal equation for the system's density operator, that takes into consideration the effects of generic distributions of time-intervals between successive resetting times. Our analysis involves paradigmatic distributions, and focuses on the examination of relevant quantities that characterise the evolution of quantum states, particularly in the study of quantum coherence and closed- vs open- system dynamics, and approaches the calculation of such quantities in the long-term regime related to non-equilibrium stationary states.

We show that stochastic resetting typically induces a loss of coherence as measured with respect to the energy eigenbasis, irrespective of the specific resetting dynamics. This allows us to identify the effects of resetting process on the evolution of the otherwise closed system, with those induced by a non-dissipative decoherence channel acting on an open system. Thus, different resetting dynamics (different distributions $\mathcal T_1(t)$) will simulate some aspects of the evolution of open systems; in particular the L\'evy-Smirnoff distribution may be of use in studying the quantum-classical transition, characterised by a complete decoherence of the initial state. 

One central aspect analysed was the loss of information on the system's state due to the renewal process. Such loss is encoded in the dynamics of the purity of the reset state, and for closed systems is a consequence of the uncertainty caused by the random restarting process. On the other hand, our analysis offers a means to quantify the \emph{effective} amount of entanglement created if the decoherence channel is ascribable to the interaction between the system and its surroundings, provided the global (composite) system is initially in a pure state.

The quantification of the distinguishability between two quantum states, as measured by the fidelity, was carried out between the reset state and its unitarily-evolving counterpart. Our analysis shows that initial pure states stand out as the states that are more affected by the resetting process, since only for them the purity and the fidelity may reach their minimum value. Further, for qubit pure states it is verified that the resetting process delays the time required for the initial state to evolve toward a distinguishable one, yet such delay can be reduced substantially by an appropriate selection of the distribution $\mathcal{T}_1(t)$ of times between consecutive reset events (in particular, in the examples analysed here the L\'evy-Smirnoff distribution reduces the delay by a factor of $\sim 10^3$).  


\ack
This work was supported by UNAM-PAPIIT IN110120 (F.J.S.) and UNAM-PAPIIT IN113720 (A.V.H).

\section*{References}

\begin{thebibliography}{10}
\expandafter\ifx\csname url\endcsname\relax
  \def\url#1{{\tt #1}}\fi
\expandafter\ifx\csname urlprefix\endcsname\relax\def\urlprefix{URL }\fi
\providecommand{\eprint}[2][]{\url{#2}}

\bibitem{EvansPRL2011}
Evans M~R and Majumdar S~N 2011 {\em Phys. Rev. Lett.\/} {\bf 106}(16) 160601
  \urlprefix\url{https://link.aps.org/doi/10.1103/PhysRevLett.106.160601}

\bibitem{FuchsEPL2016}
Fuchs J, Goldt S and Seifert U 2016 {\em {EPL} (Europhysics Letters)\/} {\bf
  113} 60009 \urlprefix\url{https://doi.org/10.1209/0295-5075/113/60009}

\bibitem{ReuveniPRL2016}
Reuveni S 2016 {\em Phys. Rev. Lett.\/} {\bf 116}(17) 170601
  \urlprefix\url{https://link.aps.org/doi/10.1103/PhysRevLett.116.170601}

\bibitem{PalPRL2017}
Pal A and Reuveni S 2017 {\em Phys. Rev. Lett.\/} {\bf 118}(3) 030603
  \urlprefix\url{https://link.aps.org/doi/10.1103/PhysRevLett.118.030603}

\bibitem{EvansJPhysA2020}
Evans M~R, Majumdar S~N and Schehr G 2020 {\em Journal of Physics A:
  Mathematical and Theoretical\/} {\bf 53} 193001
  \urlprefix\url{https://doi.org/10.1088/1751-8121/ab7cfe}

\bibitem{GuptaPRL2020}
Gupta D, Plata C~A and Pal A 2020 {\em Phys. Rev. Lett.\/} {\bf 124}(11) 110608
  \urlprefix\url{https://link.aps.org/doi/10.1103/PhysRevLett.124.110608}

\bibitem{MagoniPRR2020}
Magoni M, Majumdar S~N and Schehr G 2020 {\em Phys. Rev. Research\/} {\bf 2}(3)
  033182
  \urlprefix\url{https://link.aps.org/doi/10.1103/PhysRevResearch.2.033182}

\bibitem{RayChaos2021}
Ray A, Pal A, Ghosh D, Dana S~K and Hens C 2021 {\em Chaos: An
  Interdisciplinary Journal of Nonlinear Science\/} {\bf 31} 011103
  (\textit{Preprint} \eprint{https://doi.org/10.1063/5.0038374})
  \urlprefix\url{https://doi.org/10.1063/5.0038374}

\bibitem{SarkarChaos2022}
Sarkar M and Gupta S 2022 {\em Chaos: An Interdisciplinary Journal of Nonlinear
  Science\/} {\bf 32} 073109 (\textit{Preprint}
  \eprint{https://doi.org/10.1063/5.0090861})
  \urlprefix\url{https://doi.org/10.1063/5.0090861}

\bibitem{MukherjeePRB2018}
Mukherjee B, Sengupta K and Majumdar S~N 2018 {\em Phys. Rev. B\/} {\bf 98}(10)
  104309 \urlprefix\url{https://link.aps.org/doi/10.1103/PhysRevB.98.104309}

\bibitem{RosePRE2018}
Rose D~C, Touchette H, Lesanovsky I and Garrahan J~P 2018 {\em Phys. Rev. E\/}
  {\bf 98}(2) 022129
  \urlprefix\url{https://link.aps.org/doi/10.1103/PhysRevE.98.022129}

\bibitem{PerfettoPRB2021}
Perfetto G, Carollo F, Magoni M and Lesanovsky I 2021 {\em Phys. Rev. B\/} {\bf
  104}(18) L180302
  \urlprefix\url{https://link.aps.org/doi/10.1103/PhysRevB.104.L180302}

\bibitem{TurkeshiPRB2022}
Turkeshi X, Dalmonte M, Fazio R and Schir\`o M 2022 {\em Phys. Rev. B\/} {\bf
  105}(24) L241114
  \urlprefix\url{https://link.aps.org/doi/10.1103/PhysRevB.105.L241114}

\bibitem{Rivas2012open}
Rivas A and Huelga S~F 2012 {\em Open quantum systems\/} vol~10 (Springer)

\bibitem{NagarPRE2016}
Nagar A and Gupta S 2016 {\em Phys. Rev. E\/} {\bf 93}(6) 060102
  \urlprefix\url{https://link.aps.org/doi/10.1103/PhysRevE.93.060102}

\bibitem{ChechkinPRL2018}
Chechkin A and Sokolov I~M 2018 {\em Phys. Rev. Lett.\/} {\bf 121}(5) 050601
  \urlprefix\url{https://link.aps.org/doi/10.1103/PhysRevLett.121.050601}

\bibitem{Maso-PuigdellosasPRE2019}
Mas\'o-Puigdellosas A, Campos D and M\'endez V~m~c 2019 {\em Phys. Rev. E\/}
  {\bf 99}(1) 012141
  \urlprefix\url{https://link.aps.org/doi/10.1103/PhysRevE.99.012141}

\bibitem{MasoliverPRE2019}
Masoliver J and Montero M 2019 {\em Phys. Rev. E\/} {\bf 100}(4) 042103
  \urlprefix\url{https://link.aps.org/doi/10.1103/PhysRevE.100.042103}

\bibitem{TasakiPRL1998}
Tasaki H 1998 {\em Phys. Rev. Lett.\/} {\bf 80}(7) 1373--1376
  \urlprefix\url{https://link.aps.org/doi/10.1103/PhysRevLett.80.1373}

\bibitem{PolkovnikovRMP2011}
Polkovnikov A, Sengupta K, Silva A and Vengalattore M 2011 {\em Rev. Mod.
  Phys.\/} {\bf 83}(3) 863--883
  \urlprefix\url{https://link.aps.org/doi/10.1103/RevModPhys.83.863}

\bibitem{RadiceJPhysA2022}
Radice M 2022 {\em Journal of Physics A: Mathematical and Theoretical\/} {\bf
  55} 224002 \urlprefix\url{https://doi.org/10.1088/1751-8121/ac654f}

\bibitem{AdessoJPA2016}
Adesso G, Bromley T~R and Cianciaruso M 2016 {\em Journal of Physics A:
  Mathematical and Theoretical\/} {\bf 49} 473001
  \urlprefix\url{https://doi.org/10.1088/1751-8113/49/47/473001}

\bibitem{StreltsovRMP2017}
Streltsov A, Adesso G and Plenio M~B 2017 {\em Rev. Mod. Phys.\/} {\bf 89}(4)
  041003 \urlprefix\url{https://link.aps.org/doi/10.1103/RevModPhys.89.041003}

\bibitem{BaumgratzPRL2014}
Baumgratz T, Cramer M and Plenio M~B 2014 {\em Phys. Rev. Lett.\/} {\bf
  113}(14) 140401
  \urlprefix\url{https://link.aps.org/doi/10.1103/PhysRevLett.113.140401}

\bibitem{RadakrishnanPRA2019}
Radhakrishnan C, L\"u Z, Jing J and Byrnes T 2019 {\em Phys. Rev. A\/} {\bf
  100}(4) 042333
  \urlprefix\url{https://link.aps.org/doi/10.1103/PhysRevA.100.042333}

\bibitem{SaxenaPRR2020}
Saxena G, Chitambar E and Gour G 2020 {\em Phys. Rev. Research\/} {\bf 2}(2)
  023298
  \urlprefix\url{https://link.aps.org/doi/10.1103/PhysRevResearch.2.023298}

\bibitem{JafariPRA2020}
Jafari R and Akbari A 2020 {\em Phys. Rev. A\/} {\bf 101}(6) 062105
  \urlprefix\url{https://link.aps.org/doi/10.1103/PhysRevA.101.062105}

\bibitem{MisraJMP77}
Misra B and Sudarshan E~C~G 1977 {\em Journal of Mathematical Physics\/} {\bf
  18} 756--763 (\textit{Preprint} \eprint{https://doi.org/10.1063/1.523304})
  \urlprefix\url{https://doi.org/10.1063/1.523304}

\bibitem{ItanoJP2009}
Itano W~M 2009 {\em Journal of Physics: Conference Series\/} {\bf 196} 012018
  \urlprefix\url{https://doi.org/10.1088/1742-6596/196/1/012018}

\bibitem{LiangRPP2019}
Liang Y~C, Yeh Y~H, Mendon{\c{c}}a P~E~M~F, Teh R~Y, Reid M~D and Drummond P~D
  2019 {\em Reports on Progress in Physics\/} {\bf 82} 076001
  \urlprefix\url{https://doi.org/10.1088/1361-6633/ab1ca4}

\bibitem{LevitinPRL2009}
Levitin L~B and Toffoli T 2009 {\em Phys. Rev. Lett.\/} {\bf 103}(16) 160502
  \urlprefix\url{https://link.aps.org/doi/10.1103/PhysRevLett.103.160502}

\bibitem{DeffnerJPhysA2017}
Deffner S and Campbell S 2017 {\em Journal of Physics A: Mathematical and
  Theoretical\/} {\bf 50} 453001
  \urlprefix\url{https://doi.org/10.1088\%2F1751-8121\%2Faa86c6}

\bibitem{Mandelstam1991}
Mandelstam L and Tamm I 1991 {\em The Uncertainty Relation Between Energy and
  Time in Non-relativistic Quantum Mechanics\/} (Berlin, Heidelberg: Springer
  Berlin Heidelberg) pp 115--123 ISBN 978-3-642-74626-0
  \urlprefix\url{https://doi.org/10.1007/978-3-642-74626-0\_8}

\bibitem{MargolusPhysicaD1998}
Margolus N and Levitin L~B 1998 {\em Physica D: Nonlinear Phenomena\/} {\bf
  120} 188 -- 195 ISSN 0167-2789 proceedings of the Fourth Workshop on Physics
  and Consumption
  \urlprefix\url{http://www.sciencedirect.com/science/article/pii/S0167278998000542}

\bibitem{ValdesJPhysA2020}
Valdés-Hernández A and Sevilla F~J 2020 {\em Journal of Physics A:
  Mathematical and Theoretical\/}
  \urlprefix\url{http://iopscience.iop.org/article/10.1088/1751-8121/abcd56}

\bibitem{AolitaRPP2015}
Aolita L, de~Melo F and Davidovich L 2015 {\em Reports on Progress in
  Physics\/} {\bf 78} 042001
  \urlprefix\url{https://doi.org/10.1088/0034-4885/78/4/042001}

\bibitem{nielsen00}
Nielsen M~A and Chuang I~L 2000 {\em Quantum Computation and Quantum
  Information\/} (Cambridge University Press)

\bibitem{SallesPRA2008}
Salles A, de~Melo F, Almeida M~P, Hor-Meyll M, Walborn S~P, Souto~Ribeiro P~H
  and Davidovich L 2008 {\em Phys. Rev. A\/} {\bf 78}(2) 022322
  \urlprefix\url{https://link.aps.org/doi/10.1103/PhysRevA.78.022322}

\end{thebibliography}

\providecommand{\newblock}{}

\end{document}